\documentclass[12pt,tightenlines,eqsecnum,floats,showpacs,nofootinbib,amsmath,amssymb,aps,prd]{revtex4}
\usepackage{graphicx,verbatim}
\usepackage{amsmath}
\usepackage{amsfonts}
\usepackage{amssymb}
\usepackage{stmaryrd}
\usepackage{color}
\usepackage{float}

\def\del{\partial}
\def\grad{\nabla}
\def\bgrad{\overline{\nabla}}

\def\scri{\mathcal{I}}
\def\scrip{\mathcal{I}^{+}}
\def\rmd{\mathrm{d}}
\def\={\hat{=}}

\def\T{\mathcal{T}}
\def\scri{\mathcal{I}}
\def\Ecal{\mathcal{E}}

\def\bgrad{\overline{\nabla}}
\def\uchi{\bar{\chi}}
\def\etaR{\eta_{\text{\tiny{Ret}}}}
\def\etar{\eta_{\text{\tiny{ret}}}}
\def\b{\bar}
\def\ba{\bar{a}}

\newcommand{\pb}[1]{\hbox{\lower0.5ex\hbox{${}_{\leftarrow}$}}\kern-1.9ex{#1}}

\def\Qr{Q^{(\rho)}}
\def\Qp{Q^{(p)}}

\def\ps{\Gamma_{\!\rm Cov}}

\def\ord{\mathcal{O}}
\def\M{\Sigma}
\def\man{M^{+}_{\rm P}}

\def\t{\tilde}
\def\h{\hat}

\def\be{\begin{equation}}
\def\ee{\end{equation}}
\def\ba{\begin{eqnarray}}
\def\ea{\end{eqnarray}}

\def\f{\frac}

\def\gb{\bar{g}}
\def\bgamma{\overline{\gamma}}
\def\qo{\mathring{q}}
\def\go{\mathring{g}}
\def\gammao{\mathring{\gamma}}
\def\Do{\mathring{D}}
\def\grado{\mathring{\nabla}}

\def\boxo{\mathring{\Box}}
\def\epsilono{\mathring\epsilon}
\def\eo{\mathring{e}}
\def\etao{\mathring{\eta}}
\def\vx{\vec{x}}
\def\vy{\vec{y}}

\newcommand{\Qab}{Q^{(\rho)}_{ab}}
\newcommand{\QPab}{Q^{(p)}_{ab}}

\renewcommand{\d}{\mathrm{d}}

\begin{document}

\title{Asymptotics with a positive cosmological constant: III.\\ The quadrupole formula}
\author{Abhay Ashtekar}
\email{ashtekar@gravity.psu.edu} \affiliation{Institute for
Gravitation and the Cosmos \& Physics
  Department, Penn State, University Park, PA 16802, U.S.A.}
\author{B\'eatrice Bonga}
\email{bpb165@psu.edu} \affiliation{Institute for Gravitation and
the Cosmos \& Physics
  Department, Penn State, University Park, PA 16802, U.S.A.}
\author{Aruna Kesavan}
\email{aok5232@psu.edu} \affiliation{Institute for Gravitation and the
Cosmos \& Physics Department, Penn State, University Park, PA 16802,
U.S.A.}

\begin{abstract}

Almost a century ago, Einstein used a weak field approximation around Minkowski space-time to calculate the energy carried away by gravitational waves emitted by a time changing mass-quadrupole. However, by now there is strong observational evidence for a positive cosmological constant, $\Lambda$. To incorporate this fact, Einstein's celebrated derivation is generalized by replacing Minkowski space-time with de Sitter space-time. The investigation is motivated by the fact that, because of the significant differences between the asymptotic structures of Minkowski and de Sitter space-times, many of the standard techniques, including the standard $1/r$ expansions, can not be used for $\Lambda >0$. Furthermore since, e.g., the energy carried by gravitational waves is always positive in Minkowski space-time but can be arbitrarily negative in de Sitter space-time \emph{irrespective of how small $\Lambda$ is}, the limit $\Lambda\to 0$ can fail to be continuous. Therefore, a priori it is not clear that a small $\Lambda$ would introduce only negligible corrections to Einstein's formula. We show that, while even a tiny cosmological constant does introduce qualitatively new features, in the end, corrections to Einstein's formula are negligible for astrophysical sources currently under consideration by gravitational wave observatories.

\end{abstract}

\pacs{04.70.Bw, 04.25.dg, 04.20.Cv}

\maketitle

\section{Introduction}
\label{s1}

One of the first predictions of general relativity came from Einstein's calculations that demonstrated the existence of gravitational waves in the weak field approximation. Although the idea of gravitational waves was already explored by others including Lagrange and Poincar\'e (see \cite{Kennefick:book} for a review), Einstein's 1916 paper provided a relativistic description by linearizing field equations of general relativity off Minkowski background, in the presence of an external, time-changing source \cite{Einstein:1916}. Two years later, he also calculated the energy carried by these waves far away from the source. He found that the leading order contribution to the emitted power is proportional to the square of the third time derivative of the mass quadrupole moment \cite{Einstein:1918}. However, in the ensuing decades, there was a great deal of confusion on the question of whether gravitational waves even \emph{exist} in full general relativity, beyond the linear approximation \cite{Kennefick:art}. On the theoretical side, this controversy was finally resolved in the early 1960s by the work of Bondi, Sachs and others \cite{bondi,rp}. On the observational side, the physical reality of gravitational waves was established by the discovery of the Hulse-Taylor binary pulsar in 1974 and careful monitoring of its orbit over the subsequent years \cite{tw}. These high precision measurements allowed a direct comparison between the loss of orbital energy and the energy emitted by gravitational waves. Today, observational evidence yields a confirmation of the existence of gravitational quadrupolar radiation to an accuracy of 3 parts in $10^{3}$  
\cite{Damour:binary}.

Einstein's calculations and its subsequent refinements and generalizations (due to Eddington \cite{ase}, Landau and Lifshitz \cite{ll}, Fock \cite{vaf}, Blanchet and Damour \cite{bd} and others), as well as the Bondi-Sachs framework in full general relativity \cite{bondi,rp}, use field equations with a vanishing cosmological constant, $\Lambda$. On the other hand, by now there is strong evidence from independent observations that the dominant contribution to the energy density of the universe is best modeled by a positive cosmological constant $\Lambda$ \cite{sndata, planck}. But because  the value of $\Lambda$ is so small, at a practical level it seems natural to just ignore it and use the well-developed $\Lambda=0$ framework. However, our study of isolated gravitating systems in asymptotically de Sitter space-times in \cite{abk1} and of linear fields on a de Sitter background in \cite{abk2} showed that there are some qualitative differences between the $\Lambda=0$ and $\Lambda >0$ cases, making the limit $\Lambda \to 0$ quite subtle.%
\footnote{The origin of these subtleties lies in the fact that the observed accelerated expansion makes the asymptotic space-time geometry in the distant future drastically different from that of asymptotically Minkowski space-times. Therefore, although for concreteness and simplicity we will refer to a cosmological constant, as in \cite{abk1,abk2}, our main results will not change if instead one has an unknown form of `dark energy'.}
In particular, the limit of observable quantities associated with gravitational waves can be \emph{discontinuous}, whence smallness of $\Lambda$ does not always translate to smallness of corrections to the $\Lambda=0$ results. The question then is whether one can reliably justify one's first instinct that Einstein's $\Lambda=0$ quadrupole formula can receive only negligible corrections, given the smallness of $\Lambda$. 

To make this concern concrete, let us consider a few illustrations of the qualitative differences. First, irrespective of how small $\Lambda$ is, we do not yet have the analog of the Bondi news  tensor \cite{bondi,aa-radmodes,memory3} that describes gravitational radiation in a gauge invariant and manageable fashion in the $\Lambda=0$, non-linear general relativity \cite{rp2,abk1}. Indeed, even the radiation field $\Psi_{4}^{0}$ that is heavily used in both analytical discussions of gravitational radiation and current numerical simulations in the $\Lambda=0$ context acquires an ambiguity in the $\Lambda>0$ case called the `origin dependence' by Penrose \cite{rp,rpwr} and `direction-dependence' by Krtou\v{s} and Podolsk\'{y} \cite{kp}. Second, while wavelengths of linear fields are constant in Minkowski space-time, they increase as waves propagate on de Sitter space-time, and exceed the curvature radius in the asymptotic region of interest. Therefore, the commonly used geometric optics approximation fails in the asymptotic region. Also, one cannot carry over the standard techniques to specify `near and far wave zones' from the $\Lambda=0$ case. Third, in Minkowski space-time one can approach $\scrip$ --the arena on which properties of gravitational waves can be analyzed unambiguously-- using $r= r_{0}$ surfaces with larger and larger values of $r_{0}$. Therefore it is standard practice to use $1/r$ expansions of fields in the analysis of gravitational waves (see, e.g., \cite{pw,straumann,mtw}). By contrast, in de Sitter space-time, such time-like surfaces approach a past cosmological horizon \emph{across which there is no flux of energy or momentum for retarded solutions.} $\scrip$ is now approached by a family of space-like surfaces (on which \emph{time} is constant) whence one cannot use the $1/r$ expansions that dominate the  literature on gravitational waves. Fourth, while $\scrip$ is null in the asymptotically flat case, it is space-like if $\Lambda$ is positive \cite{rp}. Consequently, unfamiliar features can arise as we move from $\Lambda =0$ to a tiny positive value both in full general relativity and in the linearized limit. In particular, for every $\Lambda >0$, all asymptotic symmetry vector fields --including those corresponding to `time translations'--  are also space-like in a neighborhood of $\scrip$. As a result, while the energy carried by electromagnetic and linearized gravitational waves is necessarily positive in the $\Lambda\!=\!0$ case, it can be \emph{negative and of arbitrarily large magnitude} if $\Lambda >0$\, \cite{abk2}. Since this holds for every $\Lambda >0$, however tiny, the lower bound on energy carried by these waves has an \emph{infinite} discontinuity at $\Lambda=0$. Now, if (electromagnetic or) gravitational waves produced in realistic physical processes \emph{could} carry negative energy, we would be faced with a fundamental instability: the source could gain arbitrarily high energy simply by letting the emitted waves carry away negative energy. Thus a positive $\Lambda$, however small, opens up an unforeseen possibility, with potential to drastically change gravitational dynamics.%
\footnote{We suggested in \cite{abk2} that this possibility will not be realized for realistic sources because the fields they induce on $\scrip$ would be constrained just in the right way for the waves to carry positive energy. However, that argument was meant only as an indication, based on properties of source-free gravitational waves.  A detailed analysis of the quadrupole formula for $\Lambda >0$ is needed to settle this issue in the weak field limit.}
Finally, yet another difference is that in the transverse (i.e., Lorentz) traceless gauge the linearized 4-metric field satisfies the \emph{massive} Klein-Gordon equation (where the mass is proportional to $\sqrt{\Lambda}$). While the mass is tiny, a priori it is possible that over cosmological distances the difference from the propagation in the $\Lambda=0$ case could accumulate, creating an $\ord(1)$ difference in the linearized metric in the asymptotic region, far way from sources. Since Einstein's quadrupole formula is based on the form of the metric perturbation in this `wave zone', secular accumulation could then lead to $\ord(1)$ departures from  that formula, even when $\Lambda$ is tiny. 

These considerations bring out the necessity of a systematic analysis to determine whether the Einstein's quadrupole formula continues to be valid even though many of the key intermediate steps cannot be repeated for the de Sitter background. The goal of this paper is to complete this task for linearized gravitational waves created by time changing (first order) sources on de Sitter background. 

As in the $\Lambda=0$ case, the calculation involves two steps:\vskip0.1cm 
\noindent(i) expressing metric perturbations far away from the source in terms of the quadrupole moments of the source, and,\\ 
(ii) finding the energy radiated by this source in the form of gravitational waves.\vskip0.1cm 
\noindent However, the extension of the $\Lambda=0$ analysis introduces unforeseen issues in both steps. In step (i), since the background space-time is no longer flat, the meaning of `quadrupole moment' is not immediately clear. The second subtlety concerns both steps. Specifically, due to the curvature of the de Sitter space-time, the gravitational waves back-scatter. This back-scattering introduces a tail term in the solutions to the linearized Einstein's equation already in the first post-Newtonian order. That in and of itself is not problematic. However, if a tail term persisted in the formula for energy loss, one would need to know the history of the source throughout its evolution in order to determine the flux of energy emitted at any given retarded instant of time.%
\footnote{In the $\Lambda =0$ case, back-scattering occurs only at higher post-Newtonian orders. These higher order corrections to the quadrupole formula are not needed to compare theory with observations for the Hulse-Taylor pulsar so far because the current observational accuracy is at the $10^{-3}$ level rather than $10^{-4}$.}
Third, as discussed above, the energy calculated in step (ii) could, in principle, be arbitrarily negative, in which case self-gravitating systems would be drastically unstable to emission of gravitational waves.  

Thus, from a conceptual standpoint, the generalization of the quadrupole formula to include a positive $\Lambda$ is both interesting and subtle. For example, the presence of the tail term opens a door to a new contribution to the `memory effects' associated with gravitational waves \cite{memory1,memory2,memory3}. In addition, as in the asymptotically flat case, it offers guidance in the development of the full, nonlinear framework. Finally, as we will see, this generalization also provides detailed control on the approximations involved in setting $\Lambda$ to zero.  

The paper is organized as follows. In section \ref{s2}, we introduce our notation and recall the linearized Einstein's equation on de Sitter background as well as their retarded solutions sourced by a (first order) stress-energy tensor. In section \ref{s3}, we introduce the late time and post-Newtonian approximations and express the leading terms of solutions in terms of the quadrupole moments of sources. In section \ref{s4}, we use these expressions to calculate the energy emitted by the source using Hamiltonian methods on the covariant phase-space of the linearized solutions introduced in \cite{abk2}, and then discuss in some detail the novel features that arise because of the presence of a positive $\Lambda$. We find that the energy carried away by the gravitational waves produced by a time changing source is necessarily positive. Detailed expressions bear out the expectation that, for sources of gravitational radiation currently under consideration by  gravitational wave observatories, the primary modification to  Einstein's formula can be incorporated by taking into account the expansion of the universe and the resulting gravitational red-shift. Section \ref{s5} contains a brief summary. Appendix A discusses the tail term in the retarded solution which makes the limit $\Lambda\to 0$ limit quite subtle. 

We use the following conventions. Throughout we assume that the underlying space-time is 4-dimensional and set $c$=1. The space-time metric has signature -,+,+,+.  The curvature tensors are defined via:  $2\nabla_{[a}\nabla_{b]} k_c = R_{abc}{}^d k_d$, $R_{ac} = R_{abc}{}^b$ and $R = R_{ab}g^{ab}$. Throughout we use Penrose's abstract index notation \cite{rpwr,ahm}: $a,b,\ldots$ will be the abstract indices labeling tensors while indices $\b{a}, \b{b},\ldots$ will be numerical indices. In particular, components of a tensor field $T_{ab}$ (in a specified chart) are denoted by by $T_{\b{a}\b{b}}$. We have made an effort to ensure that this paper is conceptually and notationally self-contained. However, we refer the reader to our earlier papers, \cite{abk1} and \cite{abk2}, for more detailed discussions of the numerous issues raised by the inclusion of a positive cosmological constant.

\section{Preliminaries}
\label{s2}

\begin{figure}[]
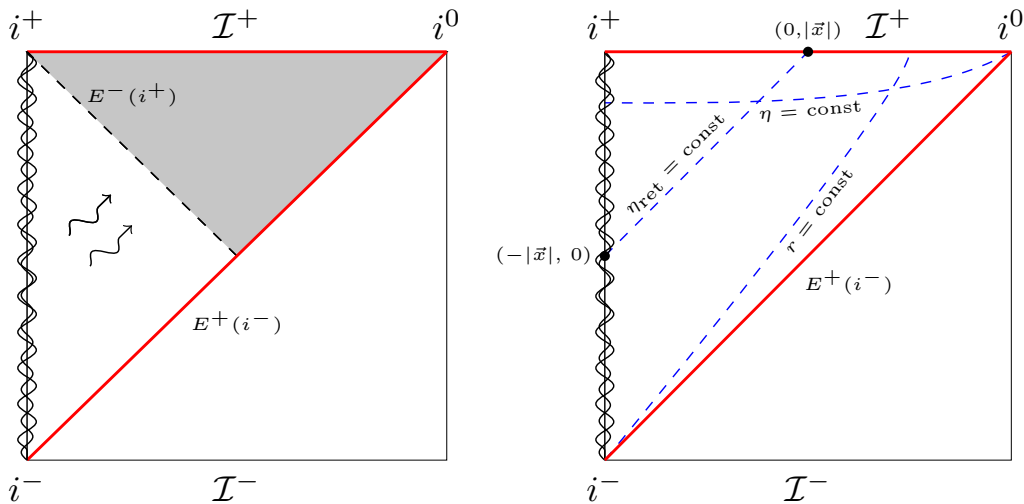

  \begin{center}
  \vskip-0.4cm
    \includegraphics[width=2.5in,height=2.7in,angle=0]{dS_shaded_braided_source.pdf}
    \includegraphics[width=2.9in,height=2.7in,angle=0]{dS_approxsolution_limit.pdf}
\caption{ {\textit{Left Panel:}} A time-changing quadrupole emitting gravitational waves whose spatial size is uniformly bounded in time. The causal future of such a source covers only the future Poincar\'e patch $\man$ (the upper triangle of the figure). There is no incoming radiation across the past boundary $E^{+}(i^{-})$ of $\man$ because we use retarded solutions.  The shaded region represents a convenient neighborhood of $\scrip$ in which perturbations satisfy a homogenous equation and the approximation (\ref{approxsoln}), discussed below, holds everywhere. The dashed (red) lines with arrows show the integral curves of the `time translation' Killing field $T^{a}$ (adapted to the rest frame of the source).\\
{\textit{Right Panel:}} The rate of change of the quadrupole moment at the point $(-|\vec{x}|, \vec{0})$ on the source creates the retarded field at the point $(0, \vec{x})$ on $\scrip$. The figure also shows the cosmological foliation $\eta= {\rm const}$ and the time-like surfaces $r:= |\vx|={\rm const}$. As $r$ goes to infinity, the $r={\rm const}$ surfaces approach $E^{+}(i^{-})$. Therefore, in contrast with the situation in Minkowski space-time, for sufficiently large values of $r$, there is no flux of energy across the $r={\rm const}$ surfaces.}
\label{poincare}
\end{center}
\end{figure}

The isolated system of interest is depicted in the left panel of Fig.~\ref{poincare} (and specified in greater detail in the beginning of section \ref{s3}). It represents a matter source in de Sitter space-time whose spatial size is uniformly bounded in time. Such a source intersects $\scri^{\pm}$ at single points $i^{\pm}$. Examples are provided by isolated stars and coalescing binary systems. The causal future of such a source covers only the future Poincar\'e patch, $\man$. No observer whose world-line is confined to the past Poincar\'e patch can see the isolated system or detect the radiation it emits. Therefore, to study this system, it suffices to restrict oneself just to $\man$, which can be coordinatized by $(\eta, x, y, z)$. On this patch, the comoving spatial coordinates, $(x,\,y,\,z)$, span the entire range $(-\infty,\,\infty)$ while the conformal time coordinate $\eta$ takes values in $(-\infty,\, 0)$. The background de Sitter metric $\bar{g}_{ab}$ takes the form 
\be
\label{dsmetric}
\gb_{ab} = a^2(\eta) \go_{ab} \quad {\rm with} \quad \go_{\b{a}\b{b}} {\rm d} x^{\b{a}} {\rm d} x^{\b{b}} = (- {\rm d} \eta^2 + {\rm d} \vx^2)\, ,\ee
%
where the scale factor is $a(\eta)= -(H\eta)^{-1}$, the Hubble parameter $H$ is related to the cosmological constant $\Lambda$ by $H:=\sqrt{\Lambda/3}$. As discussed in section 4.3.2 of \cite{abk1}, there is a seven dimensional group of isometries that leaves this metric invariant \textit{and} that maps this patch to itself. In this paper the primary focus will be the vector field generating time translations  as this is the relevant vector field to calculate energy and power. This Killing vector field is denoted by $T^a$ and is given by%
\footnote{This time translation $T^{a}$ is adapted to the rest frame of the source in the asymptotic future. In \cite{abk1} it was denoted by $D^a$ following the terminology in the literature, where its role as the dilation vector field w.r.t. $\go_{ab}$ is emphasized.}
\be \label{T} T = - H\, \left[\eta \, \f{\partial}{\partial \eta} + x \f{\partial}{\partial x} + y \f{\partial}{\partial y} + z \f{\partial}{\partial z} \right]. \ee
We will refer to it as a `time translation' because it is the limit of the time translation Killing field of the Schwarzschild-de Sitter space-time as the Schwarzschild mass goes to zero, and because it reduces to a time translation of Minkowski space-time in the limit $\Lambda$ goes to zero. The second property is not obvious from the form of the vector field above, since $T^a$ appears to vanish as $\Lambda$ (and consequently $H$) goes to zero. However, as discussed in \cite{abk2}, one has to be more careful in taking the limit $\Lambda \to 0$: One cannot use $(\eta,\, \vx)$ coordinates since the metric in \eqref{dsmetric} is not well-defined in this limit. Rather, one needs to use the differential structure induced on the Poincar\'e patch $\man$ by the $(t,\,\vx)$ coordinates, where the proper time $t$ is related to the conformal time $\eta$ via $H \eta = - e^{-H t}$. In the $(t,\,\vx)$ chart, when $\Lambda \to 0$, we have
\be
\label{metriclimit}
\gb_{\b{a}\b{b}} {\rm d} x^{\b{a}} {\rm d} x^{\b{b}} = - {\rm d} t^2 + e^{2 H t} \, {\rm d} \vx^2 \quad\to \quad - {\rm d} t^2 + {\rm d} \vx^2 =: \etao_{\b{a}\b{b}} {\rm d} x^{\b{a}} {\rm d} x^{\b{b}}
\ee
and $T^a\to t^a = \etao^{ab}\nabla_{b} t$, a time translation Killing field of the Minkowski metric $\etao_{ab}$ (which is distinct from $\go_{ab}$ which is also flat).

To study the gravitational radiation emitted by an isolated system in the presence of positive $\Lambda$, we consider first order perturbations off de Sitter space-time. The perturbed metric is denoted by $g_{ab}$,
\be
g_{ab}\, =\, \gb_{ab} + \epsilon \, \gamma_{ab}\, =:\, a^2(\eta)(\go_{ab} + \epsilon \, h_{ab})\, ,
\ee
where $\epsilon$ is a smallness parameter. While $\gamma_{ab}$ are the physical first order perturbations off de Sitter space-time, it is convenient --as will be clear shortly-- to use the conformally related mathematical field $h_{ab}$ while solving the linearized Einstein's equation. 
 
In terms of the trace-reversed metric perturbation $\overline{\gamma}_{ab} := \gamma_{ab} - \frac{1}{2} \bar{g}_{ab} \gamma$,\,\, the linearized Einstein's equation in the presence of a (first order) linearized source $T_{ab}$ can be written as
\be
\label{eom}
\overline{\Box}\bgamma_{ab} - 2 \bgrad_{(a} \bgrad^{c}\bgamma_{b)c} + \bar{g}_{ab} \bgrad^{c} \bgrad^{d} \bgamma_{cd} - \f{2}{3} \Lambda (\bgamma_{ab} - \gb_{ab}\bgamma) = - 16 \pi G \, T_{ab}
\ee
where $\bgrad$ and $\overline{\Box}$ denote the derivative operator and the d'Alembertian defined by the de Sitter metric $\bar{g}_{ab}$. 

The solutions to the linearized equation with sources on the future Poincar\'e patch $(\man, \bar{g}_{ab})$ are discussed in detail by de Vega et al. in \cite{vrs} (see also \cite{yizen} for a recent discussion). Here we will summarize their results, comment on the physical interpretation, and also discuss the limit $\Lambda \to 0$.

Denote by $\eta^{a}$ the vector field normal to the cosmological slices $\eta = {\rm const}$ satisfying $\eta^{a}\nabla_{a} \eta =1$ and let $n^a := - H \eta\, \eta^a$ denote the future pointing unit normal to these slices. Then, it is convenient to solve \eqref{eom} using the following gauge condition:
\be
\label{vrsgauge}
\bgrad^a \bgamma_{ab} = 2 H \, n^a \, \bgamma_{ab}\, .
\ee
This is a generalization of the more familiar Lorentz gauge condition and, as with the Lorentz condition, it does not exhaust the gauge freedom.  Nonetheless, in this gauge the linearized Einstein's equation \eqref{eom} simplifies significantly when it is rewritten in terms of the field $\uchi_{ab}$ which is related to the trace-reversed metric perturbations $\bgamma_{ab}$ via $\uchi_{ab}:= a^{-2} \bgamma_{ab} = h_{ab}- \f{1}{2} \go_{ab} \, \go^{cd}h_{cd}$. Finally, it is easiest to obtain solutions to \eqref{eom} by performing a decomposition of $\uchi_{ab}$ and $T_{ab}$, adapted to the cosmological $\eta = \text{const}$ slices:
\ba 
&\tilde{\chi} := (\eta^a \eta^b + \qo^{ab})\,\uchi_{ab},\qquad \chi_a := \eta^c \, \qo_a{}^{b} \, \uchi_{bc}, \qquad \chi_{ab}:=\qo_a{}^{m} \, \qo_b{}^{n} \, \uchi_{mn}, \label{decom1} \\ 
&\tilde{\T} := (\eta^a \eta^b + \qo^{ab})\,T_{ab}, \qquad \T_a := \eta^c \, \qo_a{}^{b}\, T_{bc}, \qquad \T_{ab}:=\qo_a{}^{m} \, \qo_b{}^{n}\, T_{mn}, \label{decom2}
\ea
where $\qo^{ab}$ is the (contravariant) spatial metric on a $\eta = \text{const}$ slice induced by the flat metric $\go_{ab}$, i.e.,\,\, $\qo^{ab}=\go^{ab}+ \eta^a \eta^b$. (Note that unlike $\uchi_{ab}$ in (\ref{decom1}), the stress energy tensor $T_{ab}$ in (\ref{decom2}) has neither been rescaled by $a^{-2}$ nor has it been trace-reversed.) In the chart $(\eta,\vx)$ we will use in the main body of the paper, $-(1/4H\eta)\tilde{\chi}$ is the perturbed lapse function and $(H\eta)^{-2}\, \qo^{ab} \chi_{b}$ is the perturbed shift field. Thus, as in the linearized theory off Minkowski space-time, the physical degrees of freedom associated with radiation are encoded in the totally spatial projection $\chi_{ab}$. 

It is convenient to regard the fields $\tilde{\chi}, \chi_{a}$ and $\chi_{ab}$, as living in the flat space-time $(\man, \go_{ab})$ because: (i) the gauge condition and field equations have a simple form in terms of the derivative operators defined by $\go_{ab}$; and (ii) these gauge conditions and field equations are well defined also at $\scrip$ because, as we will see in section \ref{s4}, the metric $\go_{ab}$ turns out to provide a viable conformal rescaling of $\bar{g}_{ab}$ that is well-defined at $\scrip$.  The gauge conditions (\ref{vrsgauge}) become
\be \label{gauge} \Do^{a}\chi_{ab} = \partial_{\eta}\chi_{b}\,-\, \f{2}{\eta} \chi_{b},\quad{\rm and}\quad \Do^{a}\chi_{a} = \partial_{\eta}(\tilde\chi- \chi)\,-\, \f{1}{\eta} \tilde\chi\, , \ee
where $\Do$ is the derivative operator of the spatial metric $\qo_{ab}$ and $\chi= \qo^{ab}\chi_{ab}$. In this gauge, the linearized Einstein's equation (\ref{eom}) is split into three as follows
 \ba
\label{eomcartesian1}
 &\mathring{\Box}& \Big(\frac{1}{\eta} \tilde{\chi}\Big) = - \frac{16 \pi G}{\eta}\, \tilde{\T} , \\
\label{eomcartesian2} 
 &\mathring{\Box}& \Big(\frac{1}{\eta} \chi_{a} \Big) = - \frac{16 \pi G}{\eta}\, \T_{a}, \\
\label{eomcartesian3} 
& (\mathring{\Box}& + \frac{2}{\eta} \partial_{\eta} )\, \chi_{ab}  = -16 \pi G \, \T_{ab}.
\ea
where $\mathring\Box$ is the d'Alembertian operator of the flat metric $\go_{ab}$. Using the conservation of the first order stress-energy tensor, $\bar\nabla^{a}T_{ab}=0$, it is easy to directly verify that the gauge conditions and the field equations are consistent, as they must be. 

Since we wish to impose the `no incoming radiation' boundary conditions, we will seek retarded solutions to these equations.
The first two equations, \eqref{eomcartesian1} and \eqref{eomcartesian2}, can be solved using the scalar retarded Green's function of\,\, $\mathring\Box$:
\be
\label{flatgreen} G_R^{(M)}(x,\,x') = \f{1}{4\pi |\vx - \vx'|}\,\,\, \delta(\eta - \eta' - |\vx - \vx'|)\ee
to yield
\ba \label{soln1}
&\tilde{\chi}(\eta,\,\vx)& = 4 G \, \eta \int \f{{\rm d}^3 \vx'}{|\vx - \vx'|} \; \f{1}{\etaR} \, \tilde{T}(\etaR,\, \vx'),\,\,\,{\rm{and}} \notag \\
&\chi_{\b{a}}(\eta,\,\vx) & = 4 G \, \eta \int \f{{\rm d}^3 \vx'}{|\vx - \vx'|} \; \f{1}{\etaR} \, \T_{\b{a}}(\etaR,\, \vx'),
\ea
where $\etaR$ is the retarded time related to $\eta$ and $\vx$ by $\etaR: = \eta - |\vx - \vx'|$. We could use the scalar Green's function of $\boxo$ also in the second equation because $\chi_{\b{a}}$ refer to the Cartesian components of the vector perturbation. While we will use the solutions (\ref{soln1}) in an intermediate step, the fluxes of energy, momentum and angular momentum turn out to depend only on $\chi_{ab}$ because, as we noted above, the other components correspond to linearized lapse and shift fields. 

One can  use a \emph{scalar} Green's function also for the Cartesian components of the spatial tensor field $\chi_{ab}$. However, since the operator on the left hand side of \eqref{eomcartesian3} has the extra term, $(2/\eta)\partial_{\eta}$, we cannot use the Green's function of the flat space wave operator $\mathring\Box$.  Instead, Ref. \cite{vrs} provides the retarded Green's function satisfying
\be (\mathring{\Box} + \frac{2}{\eta} \partial_{\eta} )\, G_{R}(x,x')\, = - (H^{2}\eta^{2})\, \delta (x,x'). \ee
In contrast to the flat space scalar Green's function, the solution to this equation has an additional term that extends its support to the region in which $x,x^{\prime}$ are time-like related: 
\be
\label{dsgreen}
G_R(x,x') = \f{H^2 \, \eta \, \eta'}{4\pi |\vx - \vx'|} \, \delta(\eta - \eta' - |\vx - \vx'|)
\,\,+\,\, \f{H^2}{4\pi} \, \theta(\eta - \eta' - |\vx - \vx'|)\, 
\ee
where $\theta(x)$ is the step function which is 1 when $x \geq 0$ and 0 otherwise. Therefore the solution $\chi_{ab}$ is given by
\be \label{dssoln}
\chi_{\b{a}\b{b}}(\eta, \vx) = 16\pi G \, \int {\rm d}^{3}\vx'\, \int {\rm d}\eta'\,\, G_{R}(x, x') \, \big(\f{1}{H^{2}{\eta'}^{2}}\big)\,\T_{\b{a}\b{b}} (x')\, .
\ee
To simplify the solution, one uses the identity
\ba
&\big( \f{1}{|\vx -\vx'|}\,\f{\eta}{\eta'}\big)\,\, \delta (\eta-\eta'-|\vx-\vx')\,\, +\,\, \f{1}{\eta'^{2}}\, \theta(\eta - \eta' - |\vx - \vx'|)\, \nonumber\\
&=\, \f{1}{|\vx-\vx'|}\,\delta (\eta - \eta' - |\vx - \vx'|)\,\, -\,\,\f{\partial}{\partial \eta'}\,\big(\f{1}{\eta'}\, 
\theta(\eta - \eta' - |\vx - \vx'|)\big)\, ,
\ea
in (\ref{dssoln}), integrates by parts with respect to $\eta'$, and shows that the boundary terms do not contribute for any given $(\eta, \vx)$. Then everywhere on $\man$ the solution is given 
by 
\ba
\label{soln2}
\chi_{\b{a}\b{b}}(\eta,\,\vx) &=& 4 G \int \f{{\rm d}^3 \vx'}{|\vx - \vx'|} \, \T_{\b{a}\b{b}}(\etaR,\, \vx') \notag \\
&\qquad & \qquad + \, 4 G \int {\rm d}^3 \vx' \int_{-\infty}^{\etaR} {\rm d}\eta' \, \f{1}{\eta'} \, \partial_{\eta'} \T_{\b{a}\b{b}}(\eta',\, \vx')\\ 
&\equiv& {\sharp}_{\b{a}\b{b}}\,(\eta, \vx) + {\flat}_{\b{a}\b{b}}\,(\eta, \vx)\, , \label{sharpflat} \ea
where ${\sharp}_{\b{a}\b{b}}\,(\eta, \vx)$ denotes the \emph{sharp} propagation term and ${\flat}_{\b{a}\b{b}}\,(\eta, \vx)$, the prolonged \emph{tail} term. Note that this solution relates the Cartesian components of $\chi_{ab}$ to those of $\T_{ab}$. Therefore, throughout the rest of the paper, whenever we use this solution, we will be restricting ourselves to components in the Cartesian chart. 

The retarded solution (\ref{soln2}) has an interesting feature. The first term $\sharp_{ab}$ in this expression is identical to the solution for the trace-reversed perturbations which satisfy the linearized Einstein equation (with a first order source $\T_{ab}$)  w.r.t. the Minkowski metric $\go_{ab}$. The second term $\flat_{ab}$, which is absent in the Minkowski case, depends on the entire history of the behavior of the source up to time $\etaR$. It results from the back-scattering of the perturbation by curvature in the de Sitter background. Thus, in contrast to the $\Lambda=0$ case, the propagation of the metric perturbation fails to be sharp already at the first post-Newtonian order.  The retarded solutions (\ref{soln1}) and (\ref{soln2}) satisfy the equations of motion (\ref{eomcartesian1}) - (\ref{eomcartesian3}) by construction. However, to obtain a solution to the physical problem at hand, we need to make sure that they also satisfy the gauge conditions (\ref{gauge}). One can verify that this is the case using conservation of the stress-energy tensor.

Finally, let us discuss the limit $\Lambda \to 0$. From the solution (\ref{soln2}) it is not obvious that the tail term will disappear in this limit. However, as stated above, to study this limit one needs to use the differential structure given not by the $(\eta, \vx)$ chart, but by the $(t,\,\vx)$ chart in which the de Sitter metric $\bar{g}_{ab}$ of (\ref{metriclimit}) admits a well defined limit to the Minkowski metric $\etao_{ab}$ as $\Lambda \to 0$. Using the $(t,\,\vx)$ chart, it is easy to show that the gauge condition \eqref{vrsgauge} and the linearized Einstein's equation \eqref{eom} reduce to the familiar Lorentz gauge condition and linearized Einstein's equation in Minkowski space-time, respectively,
\be
\label{flateom}
\mathring{\grad}^b \gammao_{ab} = 0, \quad {\rm and}\quad \mathring{\Box}\gammao_{ab} = - 16 \pi G \, T_{ab},
\ee
where for notational coherence the metric perturbations off the Minkowski space-time metric $\etao_{ab}$ are denoted by $\gammao_{ab}$.
Note that, while in the de Sitter case different components of the perturbation satisfy different equations, \eqref{eomcartesian1}-\eqref{eomcartesian3}, in the $\Lambda \to 0$ limit these distinct equations collapse to a single flat space scalar wave equation for all Cartesian components of $\gammao_{ab}$. Consequently, the Green's functions \eqref{flatgreen} and \eqref{dsgreen} used to solve for various components of the de Sitter perturbations,  reduce to the scalar Green's function of the flat d'Alembertian operator $\mathring\Box$ of $\etao_{ab}$ (which, as we noted before, is distinct from the flat metric $\go_{ab}$),
\be
G_R^{(M)}(x,x') = \f{1}{4\pi |\vx - \vx'|}\,\,\, \delta(t - t' - |\vx - \vx'|).
\ee
Therefore in the $(t, \vx)$ chart the retarded solutions of \eqref{flateom} are given by
\be 
\gammao_{\b{a}\b{b}}(t,\,\vx) = 4 G \int \f{{\rm d}^3 \vx'}{|\vx - \vx'|} \, T_{\b{a}\b{b}}(t - |\vx - \vx'|, \vx').
\ee
This also follows directly by first expressing the final solutions (\ref{soln1}) and (\ref{soln2}) in the $(t, \vx)$ chart and then taking the $\Lambda \to 0$ limit, as it must. Thus, our expectation that tail term should disappear in the limit $\Lambda\to 0$ is explicitly borne out.  

\section{The retarded solution and quadrupole moments}
\label{s3}

In full general relativity with positive $\Lambda$, space-times describing isolated gravitating systems are asymptotically de Sitter. To compute the energy emitted in the form of gravitational waves, one would (numerically) solve Einstein's equations by imposing an appropriate `no-incoming radiation' boundary condition, find the gravitational fields on $\scrip$, and extract the energy radiated by gravitational waves from these fields. This paper, of course, restricts itself to a simplified version of this problem using the first post-de Sitter approximation. We have already incorporated the `no incoming radiation' boundary condition through retarded Green's functions and our task is to extract physical information from the emitted gravitational waves by examining these solutions at $\scrip$. As explained in section \ref{s1}, the calculation will be performed in two steps. In the first, carried out in this section, we use physically motivated approximations to simplify the retarded solution (\ref{soln2}) in the asymptotic region near $\scrip$ and relate the leading term to the time-variation of the source quadrupole moment. The second step will be carried out in section \ref{s4}.

\subsection{The late time and post-Newtonian approximations}
\label{s3.1}

To extract physical information from Eq. (\ref{soln2}), we need to examine this solution in the asymptotic region near $\scrip$.
In linearized gravity off Minkowski space-time, one can approach $\mathring\scri^{+}$ using a family of time-like tubes $r=r_{o}$, with ever increasing values of the constant $r_{o}$. Therefore, one focuses on the form of the retarded solutions at large distances from the source, keeping the leading order $1/r$ contribution, and ignoring terms that fall-off as $1/r^{2}$. Since the conformal factor used to complete Minkowski space-time in order to attach the null boundary $\scrip$ falls-off as $1/r$, this approximation is sufficient to recover the asymptotic perturbation on $\scrip$ and extract energy, momentum and angular momentum carried by gravitational waves. In de Sitter space-time, by contrast, as mentioned in section \ref{s1}, the $r=r_{o}$ time-like surfaces approach the cosmological horizon $E^{+}(i^{-})$, rather than $\scrip$  (see Fig.~\ref{poincare}). And the flux of energy or momentum or angular momentum across $E^{+}(i^{-})$ vanishes identically for retarded solutions!  Indeed, this is precisely the `no incoming radiation condition'. (Thus, $E^{+}(i^{-})$ is analogous to $\mathring{\mathcal{I}}^{-}$ rather than $\mathring{\mathcal{I}}^{+}$ in Minkowski space-time.)  Therefore, contrary to the strong intuition derived from Minkowski space-time \cite{pw,straumann,mtw}, the $1/r$-expansions are now ill-suited to study gravitational waves. (In particular, one cannot blindly take over well-understood notions such as the `wave zone'. All these differences occur also for test electromagnetic fields on de Sitter space-time.)

As explained in \cite{abk1}, $\scrip$ of de Sitter space-time  is space-like and corresponds to the surface $\eta=0$ (see also section \ref{s4.1}). Therefore, it can be approached by a family of \emph{space-like} surfaces. The first natural candidate is provided by the cosmological slices $\eta={\rm const}$ used in section \ref{s2}. Another possibility is to use the family of space-like 3-surfaces which lie in the shaded region of the left panel of Fig.~\ref{poincare} to which $T^{a}$ and the three rotational Killing fields of $(\man, \bar{g}_{ab})$ are everywhere tangential. In this section we will use the cosmological slices and in the next section, the 3-surfaces in the shaded region. To summarize, to approach $\scrip$ and extract the radiative part of the solution, we now need a \emph{late time} approximation in place of the Minkowskian `far field' approximation.

To introduce this approximation, we first need to sharpen our restrictions on the spatial support of the matter source. These conditions will capture the idea that the system under consideration is isolated, e.g., an oscillating star or a compact binary. First, we will assume that the physical size $D(\eta)$ of the system is uniformly bounded by $D_{o}$  on all $\eta = {\rm const}$ slices.  A particular consequence of this requirement is that the source punctures $\scrip$ at a single point $i^{+}$, and $\scri^{-}$ at a single point $i^{-}$, as depicted in Fig.~\ref{poincare}. Physically, this assumption will ensure that a binary, for example, remains compact in spite of the expansion of the universe. We further sharpen the `compactness' restriction through a second requirement: $D_{o} \ll \ell_{\Lambda}$, where $\ell_{\Lambda} (= 1/H)$ is the cosmological radius.%
\footnote{Given that $\ell_{\Lambda}$ is about $5\, {\rm Gpc}$, the condition is easily met by sources of physical interest, such as an isolated oscillating star or a compact binary.}
Finally, for simplicity, we assume that the system is stationary in the distant past and distant future, i.e., $\mathcal{L}_{T} T_{ab} =0$ outside a finite $\eta$-interval. Such a system is dynamically active only for a finite time interval $(\eta_{1},\, \eta_{2})$.  This simplifying assumption can be weakened substantially to allow $\mathcal{L}_{T} T_{ab}$ to fall-off at a suitable rate in the approach to $i^{\pm}$. We use the stronger assumption just to ensure finiteness of various integrals without having to consider the fall-off conditions in detail at each intermediate step. Furthermore, given that we are primarily interested in calculating radiated power at a retarded instant of time, the assumption is not really restrictive.

With these restrictions, we can now obtain an approximate form of the solution (\ref{soln2}) which is valid near $\scrip$. Consider, then, a cosmological slice,  $\eta ={\rm const}$, and choose the Cartesian coordinates $\vx$ such that the center of mass of the source lies at the origin. The right side of (\ref{sharpflat}) expresses $\chi_{ab}$ as a sum of a sharp term and tail term. Let us first simplify the sharp term. As in the standard linearized theory off Minkowski space-time \cite{pw}, we first write it as 
\be \label{sharp1} \sharp_{\b{a}\b{b}}\,(\eta, \vx) = 4G\,\int{\rm d}^{3}x'\, \int {\rm d}^{3}y'  \, \f{\T_{\b{a}\b{b}}(\etaR,\, \vy')}{|\vx - \vx'|}\,\, \delta(\vx',\vy')\, , \ee
and Taylor-expand the $|\vx-\vx'|$ dependence of the integrand around $\vx' =0$ (recall that the integral over $\vx'$ is over a compact region around the origin, the support of $\T_{\b{a}\b{b}}$). In the Taylor expansion, each derivative $\partial/\partial x^{'\b{a}}$ can be replaced by $-\partial/\partial x^{\b{a}}$ because the $\vx'$-dependence of the integrand of the last integral comes entirely from $|\vx' - \vx|$. Hence,
\ba   
\sharp_{\b{a}\b{b}}\,(\eta, \vx) &=& \f{4G}{r}\, \Big[\int {\rm d}^{3}x'\, \Big( \T_{\b{a}\b{b}}(\etar,\, \vx') + \f{x'{}^{c}\h{r}_{c}}{r}\, \T_{\b{a}\b{b}}(\etar,\, \vx') + (x^{'c}\h{r}_{c})\, \partial_{\etar}\, \T_{\b{a}\b{b}}(\etar,\, \vx')\, + \ldots \Big)\, \Big]\nonumber\\
&=& \f{4G}{r}\, \Big[\int {\rm d}^{3}x' \T_{\b{a}\b{b}}(\etar,\, \vx') + (\f{x^{'c}_{1}\h{r}_{c}}{r})\, \int {\rm d}^{3}x'\, \T_{\b{a}\b{b}}(\etar,\, \vx') \nonumber\\
&+& (x_{2}^{'c}\,\h{r}_{c})\, \int {\rm d}^{3}x'\,\partial_{\etar}\, \T_{\b{a}\b{b}}(\etar,\, \vx')\, +\ldots \Big]\ea
where we have carried out the integral over $\vy'$ and where the $\ldots$ denote higher order terms in the Taylor expansion. Note that we have replaced 
\be  \etaR = (\eta- |\vx -\vx'|) \quad\quad {\rm by} \quad\quad 
\etar = \eta -r \ee
because the coefficients of the Taylor expansion are evaluated at $\vx' =0$. In the second step we have used the mean value theorem and  $\vx'_{1}$ and $\vx'_{2}$ are the points in the support of $\T_{\b{a}\b{b}}$, determined by this theorem. Next, using the fact that each of $|{x_{1}^{'c}\h{r}_{c}}/{r}|$ and $|{x_{2}^{'c}\h{r}_{c}}/{r}|$ is bounded by the coordinate radius of the source at $\eta=\etar$,
\be d(\etar):= D(\etar)/a(\etar)\, ,\ee
we conclude 
\be \label{sharp2} \sharp_{\b{a}\b{b}}\,(\eta, \vx) \, = \, \f{4G}{r}\, \int {\rm d}^{3}x'\, \T_{\b{a}\b{b}}(\etar,\, \vx')\, \big[\,1 + \f{\ord (d(\etar))}{r} + \f{\ord (d(\etar))}{\Delta\etar}\, \big]\, ,\ee 
where $\Delta\etar$ is the dynamical time scale (measured in the $\eta$ coordinate) in which the change in the source is of $\ord(1)$. It will be clear from section \ref{s3.2} that this is the time scale in which the change in the quadrupole moments of the source is $\ord(1)$.

Up to this point, the mathematical manipulations are essentially the same as those in the linear theory off Minkowski space-time \cite{pw}. The difference lies in the underlying assumptions and the physical meaning of the approximation scheme. A straightforward calculation relates the second and third terms in the square brackets in (\ref{sharp2}) to physical properties of the source. First, we have
\be \f{d(\etar)}{r}\, =\, \f{D(\etar)}{\ell_{\Lambda}}\, \f{(-\etar)}{r} \,\le\, \f{D_{o}}{\ell_{\Lambda}}\, (1-\f{\eta}{r})\, . \ee   
Note that to study the asymptotic form of the solution on $\scrip$, unlike in the calculation off Minkowski space-time, we cannot use a large $r$ approximation. Indeed, in the calculation of the radiated energy in \ref{s4}, we will need to integrate over a finite range 
of $r$.%
\footnote{On $\scrip$ the energy flux will be non-zero in the interval $-\eta_{2} < r< -\eta_{1}$, where $(\eta_{1}, \eta_{2})$ is the interval where the source is dynamical, i.e., $\mathcal{L}_{T} \T_{ab} \not=0$.}
While $(1-\eta/r)$ can be large,  given \emph{any} $r_{o}\not=0$, we can choose a cosmological slice $\eta={\rm const}$ sufficiently close to $\scrip$ such that for all $r>r_{o}$, $(1-\eta/r)$ is arbitrarily close to $1$, whence ${d(\etar)}/{r}$ is negligible.
This is the late-time approximation. In particular, on $\scrip$ (where $\eta=0$) we can ignore the second term in the square bracket in (\ref{sharp2}) for all $r>0$. The third term can be re-expressed as
\be \f{d(\etar)}{\Delta\etar} = \f{D(\etar)}{\Delta t_{\rm ret}} \approx v \ee
where $D$ is the physical length scale of the source and $\Delta t$  the interval in proper time in which the source changes by $\ord(1)$, and where we have used the standard reasoning from Minkowski space-time to conclude that the ratio ${D(\etar)}/{\Delta t_{\rm ret}}$ can be identified with the velocity $v$ of the source. We now use the slow motion approximation in which $v \ll 1$ (in our $c=1$ units). Thus, within our assumptions the sharp term is given by
\be \sharp_{\b{a}\b{b}}\,(\eta, \vx) \, = \, \f{4G}{r}\, \int {\rm d}^{3}x' \,\T_{\b{a}\b{b}}(\etar,\, \vx')\, .\ee
For the tail term $\flat_{\b{a}\b{b}}\,(\eta, \vx)$ in (\ref{sharpflat}), this procedure only replaces $\etaR$ by $\etar$. 

By adding the two contributions $\sharp_{\b{a}\b{b}}$ and $\flat_{\b{a}\b{b}}$, we can express $\chi_{ab}$ as follows:  %
\ba
\label{approxsoln0}
\chi_{\b{a}\b{b}}(\eta,\, \vx) \, &=&   \f{4 G}{r} \int \! {\rm d}^3 \vx' \, \, \T_{\b{a}\b{b}} (\etar, \vx') \Big[\, 1 + \ord\big(\f{D_{o}}{\ell_{\Lambda}}{\big(1-\f{\eta}{r}\big)\big)} + \ord(v)\,\Big]\;  \nonumber\\ &+& 4 G  \int_{-\infty}^{\etar} \! \! {\rm d}\eta' \, \f{1}{\eta'} \, \partial_{\eta'} \, \int \! {\rm d}^3 \vx' \, \T_{\b{a}\b{b}}(\eta', \vx')\, . \ea
(The error term arising from $\etaR \to \etar$ in the tail term is included in the square bracket in the first term.) On any $\eta=\eta_{o}$ slice, the second term in the square bracket can be neglected, in particular, for all $r> -\eta_{o}$, i.e., beyond the intersection of that slice with the cosmological horizon $E^{-}(i^{+})$. On $\scrip$, it can be neglected for all $r>0$.

Let us conclude by summarizing all the approximations that were made. First, in section \ref{s2}, we presented the retarded solution to Einstein's equations in the first post-de Sitter approximation. We then assumed that the source is compact in the sense that the physical size $D(\eta)$ of the support of the stress-energy tensor $T_{ab}$ is uniformly bounded by $D_{o}$, with $D_{o} \ll \ell_{\Lambda}$.  Finally, we used the first post-Newtonian approximation to set $v \ll 1$ in our $c=1$ units. (If one were to restore $c$, then the overall factor $4G$ would be replaced by $4G/c^{4}$ in the first term and the $\ord(v)$ term would be of 1.5 post-Newtonian order.) Note that to obtain (\ref{approxsoln}), we did not have to make any assumption relating the dynamical time scale $\Delta t_{\rm{ret}}$ of the system with the Hubble time $t_{H}=1/H$. Astrophysical sources of greatest interest to the current gravitational wave observatories satisfy $\Delta t_{\rm ret} \ll t_{H}$. We will simplify the final results using this approximation in section \ref{s4.2}. 

To avoid proliferation of symbols, from now on $\chi_{ab}(\eta, \vx)$ will stand for the approximate solution obtained by ignoring the $\ord\big((D_{o}/\ell_{\Lambda})(1-\eta/r)\big)$ and $\ord(v)$ terms relative to the $\ord(1)$ terms in (\ref{approxsoln0}). Thus, we will set
\be \label{approxsoln}
\chi_{\b{a}\b{b}}(\eta,\, \vx) \, = \, \f{4 G}{r} \int \! {\rm d}^3 \vx' \, \, \T_{\b{a}\b{b}} (\etar, \vx') + 4 G  \int_{-\infty}^{\etar} \! \! {\rm d}\eta' \, \f{1}{\eta'} \, \partial_{\eta'} \, \int \! {\rm d}^3 \vx' \, \T_{\b{a}\b{b}}(\eta', \vx')\, .\ee
and again denote the sharp and the tail terms by $\sharp_{ab}$ and $\flat_{ab}$ respectively.

\subsection{Expressing the approximate solutions in terms of quadrupole moments}\label{s3.2}

To make the relation between the energy carried by the gravitational perturbations and the behavior of the source transparent, we will now express the approximate solution in terms of multipole moments of the source. Both terms on the right side of \eqref{approxsoln} involve the integral $\int \! {\rm d}^3 \vx' \, \, \T_{\b{a}\b{b}}$ of spatial components of the stress energy tensor of the source. We can rewrite this integral in terms of time derivatives of other components, using the conservation of $T_{ab}$. Recall that this strategy is used in the $\Lambda=0$ case to express the integral entirely in terms of the second time derivative of the time-time component of $T_{ab}$, i.e., the energy density. Consequently, for perturbations off flat space, only the mass quadrupole moment is relevant in the far-field approximation. As we will now show, the situation is more complicated in the $\Lambda >0$ case because the conservation equation, $\bgrad^a T_{ab} = 0$, has additional  terms due to the expansion of the scale factor of the de Sitter background.

In the $(t,\, \vx)$ coordinates, projection of the conservation equation along $t^a$ (where, as usual, $t^{a}\partial_{a} := \partial/\partial{t})$ and $\qo_a^{\;b}$ yield, respectively, 
\begin{align}
\label{eq:conservet}
&\del_t \rho  - e^{-3 H t} \, \mathring{D}^a \T_a + H \left(3 \rho + p_1 + p_2 + p_3 \right) = 0, \\
\label{eq:conserves}
&\del_t  \T_{a}  - e^{- H t} \mathring{D}^b \T_{ab} + 2 H \, \T_{a} = 0,
\end{align}
where the matter density and pressure are defined as usual via
\be \rho = T_{ab}n^{a}n^{b} \equiv H^{2}\eta^{2}\, T_{ab}\, \eta^{a}\eta^{b}, \quad 
{\rm and} \quad p_{\b{i}} = T^{ab}\, \partial_{a} x_{\b{i}}\,\, \partial_{b}x_{\b{i}}, \ee
and where $\mathring{D}_a$ is the derivative operator compatible with the flat spatial metric $\qo_{ab}$. (In the last equation, there is no sum over $\b{i}$.) In this $(t,\,\vx)$ chart it is manifest that when $\Lambda \to 0$ (i.e., $H \to 0$), these equations reduce to the time and space projections of the conservation equation with respect to the Minkowski metric $\etao_{ab}$. Extra terms proportional to $H$ arise in de Sitter space-time due to the expansion of the scale factor. These, in particular, include all the pressure terms which appear more generally in \emph{any} spatially homogeneous and isotropic space-time. Consequently, it will turn out that  $\int \! {\rm d}^3 \vx \, \, \T_{\b{a}\b{b}}$ is related not just to the second time derivative of the mass quadrupole moment of the source as in flat space-time, but also to the analogous pressure quadrupole moment. The exact dependence on the pressure terms will be derived below. But because they are multiplied by $H$, it is already clear that these terms will have fewer time derivatives than the corresponding terms involving density. 

To recast $\int \! {\rm d}^3 \vx \, \, \T_{\b{a}\b{b}}$ in the desired form, our first task is to introduce the notion of mass and pressure quadrupole moments on the de Sitter background. Being a physical attribute of the source, the quadrupole moment at any instant of time should only depend on the physical geometry and coordinate invariant properties of the source. Therefore, we define the mass quadrupole moment as follows:
\be
\label{eq:quadr}
\Qr_{\b{a}\b{b}} (\eta) := \int_{\M}\! {\rm d}^3 V \,\, \rho(\eta) \,\, \bar{x}_{\b{a}} \, \bar{x}_{\b{b}},
\ee
where $\M$ denotes any $\eta = \text{const}$ surface with proper volume element ${\rm d}^3 V$ and $\bar{x}_{\b{a}} := a(\eta) x_{\b{a}}$ is the \emph{physical} separation of the point $\vx$ from the origin. The pressure quadrupole moment is defined similarly:
\be
\label{eq:quadp}
\Qp_{\b{a}\b{b}} (\eta) := \int_{\M}\! {\rm d}^3 V \, \, \left(p_1(\eta) + p_2(\eta) + p_3(\eta) \right) \, \bar{x}_{\b{a}} \, \bar{x}_{\b{b}} \, .
\ee
We can now use the conservation of stress-energy equations (\ref{eq:conservet}) and (\ref{eq:conserves}) to relate the integral $\int \! {\rm d}^3 \vx' \, \, \T_{\b{a}\b{b}}$ to these quadrupole moments and their time derivatives.  

This derivation follows the same steps as in the standard calculation in Minkowski space-time. We begin by noting that $\int \! {\rm d}^3 \vx' \, \, \T_{\b{a}\b{b}}(t',\,\vx') = - \int \! {\rm d}^3 \vx' \, \, (\Do^{\b{m}} \T_{\b{m}(\b{a}}) x_{\b{b})}$\, because the boundary term that arises in the integration by parts vanishes since the stress-energy tensor has compact spatial support. Using the spatial projection  \eqref{eq:conserves} of the conservation equation, we can rewrite the integral as follows:
\ba
\label{eq:tab1}
\int \! {\rm d}^3 \vx' \, \, \T_{\b{a}\b{b}}(t',\,\vx') &=& - \int \! {\rm d}^3 \vx' \, \, e^{H t'} \big(\del_{t'} + 2 H \big) \, \T_{(\b{a}} (t',\,\vx') \,\, x_{\b{b})} \notag \\
&=& \f{1}{2}\int \! {\rm d}^3 \vx' \, \, e^{H t'} (\del_{t'} + 2 H)\,\big(\Do^{\b{m}} \T_{\b{m}} (t',\,\vx')\big) \, x_{\b{a}} \, x_{\b{b}}. 
\ea
Next, we use \eqref{eq:conservet}, the projection of the conservation equation along $t^a$, to eliminate $\T_a$ in favor of the energy density and pressure: 
\be
\label{eq:tab2}
\int \! {\rm d}^3 \vx' \, \, \T_{\b{a}\b{b}}(t',\,\vx') = \f{1}{2}\int \! {\rm d}^3 \vx' \, \, e^{4 H t'} \Big[\f{\del^2 \rho}{\del t'^2} + H \f{\del}{\del t'}(\,8 \,\rho + p_1+ p_2 + p_3 ) + 5 H^2 (3 \, \rho +  p_1+ p_2 + p_3) \Big] x_{\b{a}} x_{\b{b}}.
\ee

The last step in this derivation is to express the right side of \eqref{eq:tab2} in terms of the quadrupole moments defined in \eqref{eq:quadr} and \eqref{eq:quadp}. A simple calculation yields:
\be \eo^{\b{a}}_{a}\, \eo^{\b{b}}_{b}\,\,
\int\! {\rm d}^3 \vx' \,\T_{\b{a}\b{b}}(t', \vx') = \f{1}{2 a(t')} \big[ \partial_{t'}^{2}\,\Qr_{{a}{b}} - 2 H \partial_{t'}\, \Qr_{{a}{b}} + H \partial_{t'} \Qp_{{a}{b}}](t')\, ,\ee
where $\eo^{\b{a}}_{a}$ are the basis co-vectors in the $\vx$-chart. 
Finally, using the fact that Lie derivative of any tensor field $Q_{ab}$ with respect to the time translation Killing vector field is given by $\mathcal{L}_{T} Q_{ab} = T^c \grado_c \,Q_{ab} - 2 H \,Q_{ab}$, it is straightforward to show that
\be
\label{conservation}
\eo^{\b{a}}_{a}\, \eo^{\b{b}}_{b}\,\,
\int\! {\rm d}^3 \vx' \,\T_{\b{a}\b{b}}(t', \vx') = \f{1}{2 a(t')} \big[ \mathcal{L}_{T} \mathcal{L}_{T} \Qr_{{a}{b}} + 2 H \mathcal{L}_{T} \Qr_{{a}{b}} + H \mathcal{L}_{T} \Qp_{{a}{b}} + 2 H^2 \Qp_{{a}{b}} \big](t')\, ,
\ee
Since one can readily take the limit $\Lambda \to 0$ in the $(t,\vx)$ chart, we see immediately that in this limit one recovers the familiar expression 
\be \eo^{\b{a}}_{a}\, \eo^{\b{b}}_{b}\,\,\int\! {\rm d}^3 \vx' \,\, \T_{\b{a}\b{b}}(t', \vx')\quad \to \quad \f{1}{2}\, \,[\mathcal{L}_{t} \,\mathcal{L}_{t}\, \Qr_{ab}]\ee
from the discussion of the quadrupole formula in Minkowski space-time. 

Let us return to Eq.(\ref{conservation}). Note that it  is an \emph{exact} equality within the post-de Sitter approximation; in section \ref{s3.2} we have not used the assumption $D_{o} \ll \ell_{\Lambda}$ on the size of the source, nor the post-Newtonian assumption $v \ll 1$. If we invoke, e.g., kinetic theory, then the pressure goes as $p \sim \rho v^{2}$ and can then be ignored compared to the density $\rho$. Then (\ref{conservation}) simplifies to 
\be
\label{ignorep}
\eo^{\b{a}}_{a}\, \eo^{\b{b}}_{b}\,\,
\int\! {\rm d}^3 \vx' \,\, \T_{\b{a}\b{b}}(t', \vx')\,\, \approx\,\, \f{1}{2 a(t')} \big[ \mathcal{L}_{T} \mathcal{L}_{T} \Qr_{ab}  + 2 H \mathcal{L}_{T} \Qr_{{a}{b}} + 2 H^2 \Qp_{ab} \big](t')\, .
\ee
where we have retained the last term because so far we have not made any assumption on relative magnitudes of the dynamical time scale of the system and Hubble time $1/H$. Now, in the post-Minkowski analysis, one does not have to make the assumption $p \ll \rho$ because the continuity equations (\ref{eq:conservet}) do not involve pressure terms in that case. Furthermore, in the $\Lambda >0$ case, it turns out that dropping the pressure term from the exact expression (\ref{conservation}) obscures certain conceptually important features (see footnote \ref{note}). Therefore we will retain the full expression for now.

Finally we can express the solution \eqref{approxsoln} on $(\man, \bar{g}_{ab})$ in terms of the source quadrupole moments (after a simple transformation to the $(\eta,\,\vx)$ chart). Denoting by an `overdot' the Lie derivative with respect to $T^{a}$, we obtain:
\ba
\label{approxsolnquad}
\chi_{ab}(\eta,\, \vx) &=&   \f{2 G}{r \, a(\etar)}\,\, \big[ \ddot{Q}^{(\rho)}_{ab} + 2 H \dot{Q}^{(\rho)}_{ab} + H \dot{Q}^{(p)}_{ab} + 2 H^2 Q^{(p)}_{ab} \big] (\etar) \notag \\
&+& 2 G \int_{-\infty}^{\etar} \! \! \f{{\rm d}\eta'}{\eta'} \, \partial_{\eta'} \, \f{1}{a(\eta')} \big[ \ddot{Q}^{\rho}_{ab} + 2 H \dot{Q}^{(\rho)}_{ab} + H \dot{Q}^{(p)}_{ab} + 2 H^2 {Q}^{(p)}_{ab} \big](\eta') \nonumber\\
&=:& \sharp_{ab}(\eta,\vx) \, +\, \flat_{ab}(\eta, \vx)
\ea
This expression is a good approximation to the exact solution (\ref{soln2}) everywhere on $\scri$ (except at $r=0$).

\section{Time-varying quadrupole moment and energy flux}
\label{s4}

In this section, we will carry out the second main step spelled out in section \ref{s1}: We will use the approximate solution (\ref{approxsolnquad}) to generalize Einstein's quadrupole formula for the energy $E_{T}$ carried away by gravitational waves across $\scrip$. Since linearized gravitational fields do not have a gauge invariant, local stress-energy tensor, we employ the covariant Hamiltonian framework used in \cite{abk2} to compute this energy.  

This section is divided into three parts. In the first, we will discuss the asymptotic behavior of the fields that enter the expression of  energy $E_{T}$, in the second, we will derive the quadrupole formula, and in the third, we will discuss its properties.

\subsection{$\scrip$ and the perturbed electric part $\Ecal_{ab}$ of Weyl curvature}
\label{s4.1}

As in the $\Lambda=0$ case, it is simplest to obtain manifestly gauge invariant expressions of fluxes of energy-momentum and angular momentum carried away by gravitational waves using fields defined on $\scrip$. Therefore we need to carry out a future conformal completion of the background space-time $(\man,\bar{g}_{ab})$. It is natural to seek a completion that makes $(\man,\bar{g}_{ab})$ asymptotically de Sitter in a Poincar\'e patch in the sense of \cite{abk1}.
Because the physical metric $\bar{g}_{ab}$ has the form,
\be \bar{g}_{ab} = a^{2}\, \go_{ab} \equiv (H\eta)^{-2}\, \go_{ab}\, ,\ee
it is easy to verify that such a conformal completion can be obtained by setting the conformal factor $\Omega = -H\eta$, so that the conformally rescaled 4-metric, which is smooth at $\scrip$, is simply the flat metric $\go_{ab}$. We will use this completion because all our equations in the Cartesian chart of $\go_{ab}$ and the solution $\chi_{ab}$ will then automatically hold on the conformally completed space-time, including $\scrip$. The final results, of course, will be conformally invariant as in \cite{abk1,abk2}.

The formulas for fluxes of energy-momentum and angular momentum --spelled out in sections \ref{s4.2} and \ref{s5}-- involve  the so-called \emph{perturbed electric part of the Weyl tensor,} $\Ecal_{ab}$, at $\scrip$ \cite{abk2}. Therefore, we will first express {$\Ecal_{ab}$} in terms of the metric perturbations --for which we already have the explicit expression (\ref{approxsolnquad}) in terms of the quadrupole moments-- and then discuss its properties needed in the subsequent discussion.

Recall that the local conditions included in the definition of \emph{weakly} asymptotically de Sitter space-times --and therefore satisfied  by space-times that are asymptotically Sitter in a Poincar\'e patch-- imply that the Weyl curvature of the conformally rescaled metric must vanish at $\scri$ and therefore $\Omega^{-1} C_{abc}{}^{d}$ admits a smooth limit there \cite{rp,abk1}. Our conformally rescaled metric $\go_{ab}$ is flat, whence the limit of $\Omega^{-1} \mathring{C}_{abc}{}^{d}$ also vanishes. Therefore, not only is the first order perturbation ${}^{(1)}\mathcal{C}_{abc}{}^{d}$ such that $\Omega^{-1} \, ({}^{(1)}\mathcal{C}_{abc}{}^{d})$ admits a limit to $\scrip$, but furthermore the limit is \emph{gauge invariant.} The field of interest is the limit to $\scrip$ of its electric part,
\be \Ecal_{ab} :=  \Omega^{-1} \, ({}^{(1)}\mathcal{C}_{amb}{}^{n}\, \eta^{m}\eta_{n}) \,=\,  -\, (H\eta)^{-1}\, ({}^{(1)}\mathcal{C}_{amb}{}^{n}\,\eta^{m}\eta_{n})\, , \ee
where, as before, $\eta^{a}$ is the unit normal to the cosmological slices ($\eta= {\rm const}$) w.r.t. the conformal metric $\go_{ab}$ and the indices are raised and lowered also using $\go_{ab}$. We need to express $\Ecal_{ab}$ in terms of the (trace-reversed, rescaled) metric perturbation $\bar{\chi}_{ab}$ produced by the source.  This can be accomplished using the expression of ${}^{(1)}\mathcal{C}_{abc}{}^{d}$ in terms of the metric perturbation $\bar{\gamma}_{ab}$, and the equation of motion \eqref{eomcartesian3}.
The final result is:
\be \label{Ecal}
\Ecal_{ab}\,=\, \f{1}{2 H \eta}\,\, (\qo_a{}^c \qo_b{}^d - \f{1}{3} \qo_{ab} \qo^{cd})\,\,\big[\f{1}{2}\Do_c \Do_d \tilde{\chi} - \Do_{(c} \Do^m \chi_{d)m} -  \Do_{(c}\, \partial_{\eta}\chi_{d)} + (\del_\eta^2 - \f{1}{\eta} \del_{\eta}) \chi_{cd} \big].
\ee
Let us discuss the limit of each term to $\scrip$. Although we already know from general considerations that the left side of (\ref{Ecal}) admits a smooth limit to $\scrip$, some care is needed
to evaluate the right hand side because there is a $(1/\eta)$ pre-factor, and $\eta=0$ at $\scrip$. However, because the explicit retarded solutions \eqref{soln1} decay as $\eta$, one can show that the terms involving $\t\chi$ and $\chi_{\b{a}}$ admit a smooth limit to $\scri$. A more detailed calculation using (\ref{approxsoln}) shows that the fourth term, $({1}/{\eta}) \big(\del_\eta^2 - \f{1}{\eta} \del_{\eta} \big) \chi_{ab}$, also has a smooth limit to $\scrip$:
\be \label{source}
\frac{1}{\eta} \left( \del_{\eta}^2 - \frac{1}{\eta} \del_{\eta} \right) \chi_{\b{a}\b{b}}\, =\, \frac{4G}{r} \Big[ \frac{1}{\etar } \del^2_{\eta} \int \! \d^3 \vx' \, \T_{\b{a}\b{b}}(\etar,\vx') - \frac{1}{\etar^2} \del_{\eta} \int \! \d^3 \vx' \, \T_{\b{a}\b{b}}(\etar, \vx') \Big]\, .
\ee
%
%
%
Thus, we have expressed $\Ecal_{ab}$ at $\scrip$ in terms of the perturbed metric, as required. In particular, in spite of the presence of a $(1/\eta)$-pre-factor in (\ref{Ecal}), each of the four terms in that formula for $\Ecal_{ab}$ has well-defined limits to $\scrip$.
Note, incidentally, that in this calculation not only does the tail term $\flat_{ab}$ in $\chi_{ab}$ contribute but the result would diverge at $\eta=0$ without it. However, the process of taking derivatives has made the integral over $\eta'$ in $\flat_{ab}$ disappear, showing that the propagation of the left side of (\ref{source}) sharp. These features and Eq. (\ref{source}) in particular will play an important role in section \ref{s4.2}.

We will now discuss the properties of $\Ecal_{ab}$ that will be needed in subsequent calculations. First, the field equations satisfied by the first order perturbation ${}^{(1)}C_{abc}{}^{d}$ are conformally invariant. Since they are completely equivalent to the field equations satisfied by the first order Weyl tensor in the flat space-time $(\man, \go_{ab})$, we know that the propagation of\,\, ${}^{(1)}C_{abc}{}^{d}$ is sharp along the null cones of $\go_{ab}$ (which are the same as the null cones of the de Sitter metric $\bar{g}_{ab}$). Therefore the expression of the field $\Ecal_{ab}$ at $\scrip$ in terms of source quadrupole moments cannot have any tail terms. Indeed, one can verify this explicitly using the expression (\ref{Ecal}) and the exact solutions (\ref{soln1}) and (\ref{soln2}). Second, in any neighborhood of $\scrip$ where there are no matter sources, the field $\Ecal_{ab}$ is divergence-free
\be  \Do^{a} \Ecal_{ab}\, =\, 0.  \ee
Thus, $\Ecal_{ab}$ is transverse, traceless on $\scrip$. This property will make the gauge invariance of our expression of energy flux  transparent. 

Finally, as one would expect from the fact that $\Ecal_{ab}$ is gauge invariant, only the transverse-traceless (TT) components of $\chi_{ab}$ (in its decomposition into irreducible parts) contribute to $\Ecal_{ab}$. Let us begin with a standard decomposition of the ten components of the (rescaled, trace-reversed) metric perturbation $\bar{\chi}_{ab}$:
\ba \label{decomposition}
\tilde{\chi} &:=&(\eta^a \eta^b + \qo^{ab})\,\uchi_{ab}\, ,  \qquad  \qquad  \chi := \qo^{ab} \chi_{ab}, \qquad  \qquad \chi_a =: \Do_a A + A_a^T, \notag \\
\chi_{ab} &=:& \f{1}{3} \qo_{ab} \, \qo^{cd} \chi_{cd} + \left(\Do_a \Do_b - \f{1}{3} \qo_{ab} \Do^2 \right) B + 2 \Do_{(a} B_{b)}^T + \chi_{ab}^{TT},  \label{eq:decomposition}
\ea
where $A_{a}^{T}$ and $B_{a}^{T}$ are transverse and $\chi_{ab}^{TT}$ is transverse, trace-less,
\be
\Do^a A_a^T = 0	\qquad \Do^a B_a^T = 0 \qquad \Do^a \chi_{ab}^{TT} =0 \qquad \qo^{ab} \chi_{ab}^{TT} =0\, ,
\ee
and $\tilde{\chi}$, $\chi$, $B$, $\Do_{a}A$ are the longitudinal modes. Using the gauge condition \eqref{vrsgauge}, one can show that, in the expression (\ref{Ecal}) of $\Ecal_{ab}$, all contributions from the longitudinal and trace parts of $\uchi_{ab}$ cancel out and $\Ecal_{ab}$ depends only on $\chi_{ab}^{TT}$:
\be \label{Ecal2}
\Ecal_{ab}\, = \, \,\,\f{1}{2 H \eta}\,\,\big[\del_\eta^2 - \f{1}{\eta} \del_{\eta}\big] \chi_{ab}^{TT} \, ,
\ee
Since $\Ecal_{ab}$ and $\chi_{ab}^{TT}$  are both gauge invariant, the final relation (\ref{Ecal2}) holds in any gauge. The limit to $\scrip$ of this equality will play an important role in the next two sub-sections.\\ 

\emph{Remark:} In the literature on gravitational perturbations off Minkowski space-time, there is often confusion regarding the decomposition of spatial, symmetric tensors such as $\chi_{ab}$ into its irreducible parts. While studying vacuum solutions to linearized Einstein's equations, one generally uses the notion spelled out in Eq. (\ref{decomposition}) (see e.g. Box 5.7 in \cite{pw}, or section 4.3 in \cite{straumann}, or section 35.4 of \cite{mtw}). In particular, by $\chi_{ab}^{TT}$ one means  the trace-free and divergence-free part of (the spatial tensor) $\chi_{ab}$ as in (\ref{decomposition}). This usage is standard in cosmology, e.g. in the presentation of results by BICEP and Planck collaborations. It is also used heavily in the (perturbative) quantum gravity literature; for example, the conclusion that the graviton has spin 2 is arrived at by calculating the Casimir operators of the Poincar\'e group on the 1-graviton Hilbert space constructed from the Minkowski space-time analog of $\chi_{ab}^{TT}$.

But then in the study of retarded fields produced by compact sources, one uses an entirely different decomposition: Here, the $1/r$-part of $\chi_{ab}$ (i.e., the far field approximation) of the full retarded solution is projected into radial and the orthogonal spherical directions in \emph{physical space}. Unfortunately, these projections are also referred to as the trace, longitudinal and transverse-traceless parts of $\chi_{ab}$. For concreteness, let us denote by $P_{a}^{c}$ the projection operator in to the 2-sphere orthogonal to the radial direction in the physical space and set $\chi_{ab}^{tt} = (P_{a}{}^{c} P_{b}{}^{d} - (1/2) P_{ab} P^{cd})\chi_{cd} $. In the literature, in place of $tt$, the symbol $TT$ is used also for this projection (see, e.g., chapter 11 of \cite{pw}, or section 4.5.1 in \cite{straumann}, or section 36.10 in \cite{mtw}). This is confusing because the two notions of transverse traceless parts are distinct and inequivalent. The first notion is local in momentum space and the resulting $\chi_{ab}^{TT}$ is exactly gauge invariant everywhere in space-time. The second notion, which we will continue to denote by $\chi_{ab}^{tt}$, is local in the physical space and $\chi_{ab}^{tt}$ is gauge invariant only in a weaker sense involving $1/r$ fall-offs. Nonetheless, it is $\chi_{ab}^{tt}$ that is well-tailored to the Bondi-Sachs formalism at null infinity of asymptotically flat space-times.

As we have seen in section \ref{s3.1}, the $1/r$-expansion is not very useful at $\scrip$ of de Sitter space-time. Therefore in the $\Lambda >0$ discussion we only use the first decomposition, spelled out explicitly in (\ref{decomposition}).  We will refer to the second notion only in the discussion of the $\Lambda \to 0$ limit.

\subsection{Fluxes across $\scrip$}
\label{s4.2}

Let us calculate the flux of energy associated with the time translation $T^{a}$ across $\scrip$. Since $T^{a}$ is a Killing field of the background space-time $(\man, \bar{g}_{ab})$ we know that, for any choice of admissible conformal completion, $T^{a}$ admits a smooth extension which is tangential to $\scrip$. For the choice $\Omega = -H\eta$ of the conformal factor we made above, $T^{a}$ also serves as the 
dilation w.r.t. the intrinsic 3-metric $\qo_{ab}$ on $\scrip$:
\be T\,\, \hat{=} \,\, - H\, \big[ x \f{\partial}{\partial x} + y \f{\partial}{\partial y} + z \f{\partial}{\partial z} \big]. \ee
From the detailed analysis of the covariant phase space $\ps$ carried out in \cite{abk2}, the total energy flux $E_{T}$ across $\scrip$ is given by the Hamiltonian generating the time translation $T^{a}$ on $\ps$. The result can be expressed most simply in terms of $\Ecal_{ab}$ and the Lie derivative of the metric perturbation w.r.t. $T^{a}$ at $\scrip$:
\be \label{H3} E_{T}\,\, \hat{=}\,\,  \f{1}{16\pi GH}\, \int_{\scrip}\!\! \! \rmd^{3}x\,\, \Ecal_{cd}\,\big(\mathcal{L}_{T} \chi_{ab} +2H\,\chi_{ab}\big)\, \qo^{ac}\qo^{bd}\, . 
\ee
Note that because $\Ecal_{ab}$ is transverse-traceless ($TT$), the integral automatically extracts the $TT$ part of the term in the bracket and we have
\ba \label{H4} E_{T}\,\, &\hat{=}&\,\, \f{1}{16\pi G H}\,\int_{\scrip}\!\rmd^{3}x\,\, \Ecal_{cd}\,\big(\mathcal{L}_{T} \chi_{ab} +2H\,\chi_{ab}\big)^{TT}\, \qo^{ac}\qo^{bd}\, \nonumber\\
&\hat{=}& \,\, \f{1}{16\pi G H}\,\int_{\scrip}\! \rmd^{3}x\,\, \Ecal_{cd}\,\big(T^{m}\mathring{\nabla}_{m}\chi_{ab}\big)^{TT}\,\qo^{ac}\qo^{bd}\, ,
\ea
where in the second step we have used the fact that  $(\mathcal{L}_{T} \chi_{ab} + 2H\, \chi_{ab}) = T^{m}\mathring{\nabla}_{m} \chi_{ab}$.
We note on the side that, because the derivative $T^{m}\mathring\nabla_{m}$ commutes with the operation of taking the $TT$ part on $\scrip$, the integral can be rewritten as 
\be E_{T}\,\,\hat{=}\,\, \f{1}{16\pi G H}\,\int_{\scrip}\! \rmd^{3}x\,\, \Ecal_{cd}\,\big(T^{m}\mathring{\nabla}_{m}\chi_{ab}^{TT}\big)\,\qo^{ac}\qo^{bd}\, \ee
which is manifestly gauge invariant.

Next, we return to (\ref{H4}) and use (\ref{Ecal2}) to express $\Ecal_{ab}$ in terms of the $TT$-part of $\chi_{ab}$. Using that fact that the operator $(1/\eta)[\del_\eta^2 - \f{1}{\eta} \del_{\eta}\big]$ commutes with the operation of taking the TT part, we have:
\ba \label{H5}
E_T \,\,&\hat{=}&\,\, \lim_{\to \scri}\,\, \f{1}{32 \pi G H^{2}} \int \! {\rm d}^3x \, \Big[\f{1}{\eta}\,\big(\del_\eta^2 - \frac{1}{\eta} \del_\eta \big) \chi_{ab}\Big]^{TT} \, \Big[T^m \mathring{\nabla}_m \chi_{cd}\Big]^{TT} \,\,\qo^{ac} \qo^{bd}\, \nonumber\\ 
&\hat{=}& \lim_{\to \scri}\,\, \f{1}{32 \pi G H^{2}} \int \! {\rm d}^3x \, \Big[\f{1}{\eta}\,\big(\del_\eta^2 - \frac{1}{\eta} \del_\eta \big) \chi_{ab}\Big] \, \Big[T^m \mathring{\nabla}_m \chi_{cd}\Big]^{TT}\,\,\qo^{ac} \qo^{bd}\
\ea
where in the second step we removed the $TT$ on the first square bracket because the second square bracket is already $TT$ and therefore the integral automatically extracts only the $TT$ part of the first square bracket. These expressions hold for any solution $\chi_{ab}$ that is source-free in a neighborhood of $\scrip$ (e.g. within the shaded region in the left panel of Fig.~\ref{poincare}).

We now use the approximations $D_{o}/\ell_{\Lambda} \ll 1$ and $v \ll 1$ spelled out in section \ref{s3.1} and insert in (\ref{H5}) the convenient expression of $\chi_{ab}$ given in (\ref{approxsoln}). For the first square bracket we use (\ref{source}) and $\partial_{\eta} f(\eta - r) = - \partial_{r} f(\eta - r)$ and evaluate the expression at $\scrip$ by setting $\eta=0$. The result is:
\be \label{source2}
\f{1}{\eta}\,\, \Big[\big(\del_\eta^2 - \frac{1}{\eta} \del_\eta \big) \chi_{\b{a}\b{b}}\Big]\,(\vx)\,\, \hat{=}\, -\f{4G}{r}\, \partial_{r}\, \left( \f{1}{r}\, \partial_{r}\, \int {\rm d}^{3}x'\, \T_{\b{a}\b{b}}(\etar, \vx')\right). \ee
As we noted after (\ref{source}), although the tail term $\flat_{ab}$ 
in the expression (\ref{sharpflat}) of $\chi_{ab}$ does contribute to the result, the process of taking derivatives has made the integral over $\eta$ in $\flat_{ab}$ disappear and the result depends only on what the source does at time $\eta=\etar$

Next, consider the second square bracket in the integrand of (\ref{H5}). Since the term multiplying this bracket has a well-defined limit to $\scrip$, we can replace $T^{m}$ by its limiting value $-H r \h{r}^{m}$ at $\scrip$. Using (\ref{approxsoln}) we again find that, although the tail term $\flat_{ab}$ does contribute to the result, the integration over $\eta$ disappears because of the directional derivative along $T^{a}$ and we obtain
\be \label{radial} 
\Big[T^m \mathring{\nabla}_m \chi_{\b{c}\b{d}}\Big](\vx)\, =\, \f{4GH}{r}\,\int{\rm d}^{3}x'\, \T_{\b{a}\b{b}}(\etar, \vx')\, . \ee
Substituting (\ref{source2}) and (\ref{radial}) in (\ref{H5}), performing an integration by parts, and using Eq.~\eqref{conservation} to express the integral over the stress-energy tensor in terms of quadrupole moments, we obtain
\ba E_T \,\,\hat{=}\,\, \f{G}{8\pi H}\, \int \f{\rm{d}r}{r}\, {\rm d}^{2}S\!\!\! &&\Big[\, \Big(\partial_{r}\, Hr\big(\ddot{Q}_{ab}^{(\rho)}+2H\, \dot{Q}_{ab}^{(\rho)}+ H \dot{Q}_{ab}^{(p)} + 2H^{2}{Q}_{ab}^{(p)}\big)\,\Big)\times \nonumber\\
&&\Big(\partial_{r}\, Hr\big(\ddot{Q}_{cd}^{(\rho)}+2H\, \dot{Q}_{cd}^{(\rho)}+ H \dot{Q}_{cd}^{(p)} + 2H^{2}{Q}_{cd}^{(p)}\big)^{TT}\Big)\,\Big]\, \qo^{ac} \qo^{bd},\ea
where ${\rm d}^{2}S$ is the unit 2-sphere volume element of the flat metric $\qo_{ab}$ at $\scrip$, and, as before an `overdot' denotes the Lie derivative w.r.t. $T^{a}$. Finally, using the fact that the operation $r\partial_{r}$ commutes with the operation of extracting the $TT$ part and that the affine parameter $T$ along the integral curves of $T^{a}$ satisfies ${\rm d} T = {\rm d}r/(rH)$ at $\scrip$, we obtain
\be
\label{energy}
E_T\, \hat{=}\, \frac{G}{8 \pi} \int_{\scrip}\!\! {\rm d} T\, {\rm d}^{2}S\, \Big[\mathcal{R}_{ab}(\vx)\, \mathcal{R}_{cd}^{TT}(\vx)\, \qo^{ac}\qo^{bd}\,\Big]\, ,\ee
where the `radiation field' $\mathcal{R}_{ab}(\vx)$ on $\scrip$ is given by 
\be \label{R} \mathcal{R}_{ab}(\vx) \hat{=} \Big[\dddot{Q}_{ab}^{(\rho)} +3 H \ddot{Q}_{ab}^{(\rho)} + 2H^2 \dot{Q}^{(\rho)}_{ab} + H  \ddot{Q}_{ab}^{(p)} + 3H^2  \dot{Q}_{ab}^{(p)} + 2 H^3 \Qp_{ab}\Big](\etar)\, , \ee
where, as before, $\etar = \eta-r \,\,\hat{=}\,\, -r$. Note that $\mathcal{R}_{ab}$ is a \emph{field} on $\scrip$ because, given a point $\vx$ on $\scrip$, the quadrupole moments ${Q}_{ab}^{(\rho)}$ and ${Q}_{ab}^{(p)}$ are obtained by performing an integral over the source along the 3-surface  $\eta = \etar$ and these 3-surfaces change as we change $\vx$ on $\scrip$ (see Fig.~\ref{poincare}). This occurs also in the standard quadrupole formula in flat space. There is, however one difference from the standard formula: (\ref{energy}) uses the $TT$ decomposition rather than the $tt$ decomposition. (Indeed, since the $tt$ decomposition used in the flat space analysis is tied to the $1/r$-expansion, it is not very useful in the de Sitter context.) One consequence is that the $TT$ label appears only on the $\mathcal{R}_{cd}$ term in (\ref{energy}); the term $\mathcal{R}_{ab}$ is not automatically $TT$ because the volume element in (\ref{energy}) is not ${\rm d}^{3}x$. Finally, while components of individual terms such as $\dddot{Q}_{\b{a}\b{b}}^{(\rho)}(0,\vx)$ depend only on $r\equiv |\vx|$ \,at $\scrip$ and not on angles, an angular dependence is introduced while taking the $TT$ part. Therefore, the total integrand of (\ref{energy}) has a genuine angular dependence; otherwise one could have trivially performed the angular integral and replaced it just by a $4\pi$ factor. Again, conceptually, this situation is the same as for the standard quadrupole formula in flat space-time the $tt$ operation also introduces angular dependence.

Finally, as in the $\Lambda=0$ calculation, let us extract power $P_{T}$ radiated by the system at any `instant of time' $T_{0}$ at $\scrip$ 
(i.e., a 2-sphere cross-section of $\scrip$, orthogonal to the orbits of the `time-translation' $T^{a}$):
\be \label{power}
P_T (T_{0})\,\hat{=}\, \frac{G}{8 \pi} \int_{T=T_{0}}\!\! {\rm d}^{2}S\,\Big[\mathcal{R}^{ab}(\vx)\, \mathcal{R}_{ab}^{TT}(\vx)\,\Big]\ee
While the expression (\ref{H3}) of radiated energy is completely local in $\chi_{ab}$ a degree of non-locality enters while casting it in terms of sources: (\ref{power}) involves only the $TT$-part of one of the `radiation fields'. However, because the $TT$-part is taken only for one of the two `radiation fields', one can show that if $\mathcal{L}_{T} T_{ab} =0$ at an instant $\eta_{o}$, then the power at $\scrip$ vanishes at the cross-section $T=T_{0}$ representing the intersection of $\scrip$ with the null cone with vertex $(\eta_{o}, \vx= {\vec{0}})$.     

The expression (\ref{energy}) of radiated energy is the main result of this section. As in Einstein's quadrupole formula, it has been derived using the first post-Newtonian approximation under the assumption that we have an externally specified, first order stress-energy tensor $T_{ab}$ satisfying the conservation equation with respect to the background metric. \\

{\emph{Remark:} The covariant phase space $\ps$ constructed and used in \cite{abk2} to obtain flux formulas at $\scrip$ consists of \emph{homogeneous} solutions to linearized Einstein's equations. In this paper, we are considering retarded solutions with a first order source $T_{ab}$. However, in the shaded neighborhood of $\scrip$ shown in the left panel of Fig. 1, all (trace-reversed) metric perturbations $\bar{\gamma}_{ab}$  satisfy the homogeneous equation and there is a family if Cauchy surfaces for this neighborhood that approach $\scrip$. Therefore, one can use the covariant phase space framework in this neighborhood to calculate fluxes of energy, momentum and angular momentum carried by the perturbations $\bar{\gamma}_{ab}$ across $\scrip$. In this calculation, we used the leading-order terms in the expression (\ref{approxsolnquad}) of $\chi_{ab}$, ignoring terms of order $\ord\big((D_{o}/\ell_{\Lambda})(1-\eta/r)\big)$ and $\ord(v)$ compared to terms of order $\ord(1)$. However, as noted above, the simplified formula (\ref{approxsoln}) for $\chi_{ab}$ is valid in an entire neighborhood os $\scrip$ (the shaded region in the left panel of Fig.~\ref{poincare}). Finally note that, since the flux formula is gauge invariant, the calculation can be carried out in any gauge.}

\subsection{Properties of fluxes across $\scrip$}
\label{s4.3}

Our formula of the energy carried by gravitational waves across $\scrip$ have several interesting features which we now discuss in some detail.\\

(1) First, the cosmological constant term does survive (through $H = \sqrt{\Lambda/3}$) even at $\scrip$. Nonetheless, we explicitly see that, in this first post-Newtonian approximation, the radiated energy is still quadrupolar.\\ 

(2) As we discussed in section \ref{s4.1}, because of its conformal properties, it is clear that $\Ecal_{ab}$ has sharp propagation. However, the fundamental formula (\ref{H3}) for the energy flux, we started out with depends also on $\chi_{ab}$ whose expression does contain an integral over all $\eta'$ that extends all the way back to $\eta'= -\infty$. So, why is there no such integral in the final expressions of radiated energy? The reason is that what features in  (\ref{H3}) is not $\chi_{ab}$ itself but rather, \emph{its derivative,} $(\mathcal{L}_{T}+ 2 H) \chi_{ab} \,\hat{=}\, - H r \,\partial_r \chi_{ab}$. The integral over $\eta'$  disappears while taking this derivative, as we saw in (\ref{radial}). This is why our quadrupole formula (\ref{energy}) does not contain an explicit tail term in spite of back-scattering due to the background de Sitter curvature. As in the asymptotically flat case, of course, tail terms will arise in higher post-Newtonian orders.\\

(3) In contrast to the Einstein formula, there is a contribution from the time variation of the \emph{pressure} quadrupole and, furthermore, from the \emph{pressure quadrupole itself.} It is well known from the Raychaudhuri equation in cosmology that pressure contributes to gravitational attraction in any Friedmann, Lema\^{i}tre, Robertson, Walker universe. Eq.~(\ref{energy}) shows that, if $\Lambda >0$, it also \emph{sources} gravitational waves already in the leading order post-Newtonian approximation. If $p \ll \rho$ (in the $c=1$ units) as for Newtonian fluids, then the pressure terms $H\ddot{Q}_{ab}^{(p)}+ 3H^{2}\dot{Q}_{ab}^{(p)}$ can be neglected compared to the density terms $3H \ddot{Q}_{ab}^{(\rho)}+ 2H^{2}\dot{Q}_{ab}^{(\rho)}$ and the expression (\ref{R}) of $\mathcal{R}_{ab}$ simplifies to:
\be \mathcal{R}_{ab}(\vx) = \Big[\dddot{Q}_{ab}^{(\rho)} +3 H \ddot{Q}_{ab}^{(\rho)} + 2H^2 \dot{Q}^{(\rho)}_{ab} + 2 H^3 \Qp_{ab}\Big](\etar)\, . \ee
For compact binaries of immediate interest to the gravitational wave detectors, we also have $(\Delta t_{\rm ret})/t_{H} \ll 1$ where $\Delta t_{\rm{ret}}$ is the dynamical time scale in which the mass and pressure  quadrupole change by factors of $\ord(1)$ and $t_{H}$, the Hubble scale.%
\footnote{This need not be the case for the \emph{very} long wave length emission due to the coalescence of supermassive black holes.}
Then the formula further simplifies and acquires a form similar to that of the $\Lambda=0$ Einstein formula:
\be \mathcal{R}_{ab}(\vx) = \dddot{Q}_{ab}^{(\rho)}(\etar) \ee
When $\Lambda$ is as tiny as the observations imply, the de Sitter quadrupole and its `overdots' are extremely well approximated by those in Minkowski space-time and the $\Lambda >0$ first post-Newtonian approximation is extremely well-approximated by the standard one. The full expression (\ref{R}) provides a precise control over the errors one makes while using the Einstein formula in presence of $\Lambda$.\\

(4) Positivity of energy flux is not transparent because the integrand of (\ref{energy}) is not manifestly positive, as it is in Einstein's formula for flat space. However, one can establish positivity as follows. First, properties of the retarded Green's function imply that the $\chi^{TT}_{ab} (\eta, \vx)$ can be expressed using the $TT$ part $\T^{TT'}_{ab}$ of $\T_{ab}(\eta,\vx')$, where the prime in $TT'$ emphasizes that the transverse traceless part refers to the argument $\vx'$:
\be 
\label{solnTT}
\chi_{\b{a}\b{b}}^{TT}(\eta,\,\vx) = 4 G \int \f{{\rm d}^3 \vx'}{|\vx - \vx'|} \, \T^{TT'}_{\b{a}\b{b}}(\etaR,\, \vx') 
+ \, 4 G \int {\rm d}^3 \vx' \int_{-\infty}^{\etaR} {\rm d}\eta' \, \f{1}{\eta'} \, \partial_{\eta'} \T^{TT'}_{\b{a}\b{b}}(\eta',\, \vx')\, .\ee
(The $TT$ in $\chi_{\b{a}\b{b}}^{TT}(\eta,\,\vx)$ on the left side refers to $\vx$.) Next, let us rewrite the expression (\ref{H5}) in terms of $\chi^{TT}_{ab}$ 
\be \label{HTT}
E_T \,\,\hat{=}\,\, \lim_{\to \scri}\,\, \f{1}{32 \pi G H^{2}} \int \! {\rm d}^3x \, \Big[\f{1}{\eta}\,\big(\del_\eta^2 - \frac{1}{\eta} \del_\eta \big) \chi_{ab}^{TT}\Big] \, \Big[T^m \mathring{\nabla}_m \chi_{cd}^{TT}\Big] \,\,\qo^{ac} \qo^{bd}\, \ee
where we have used the fact that $\partial_{\eta}$ and $T^m \mathring{\nabla}_m$ commute with the operation of taking the $TT$ part. Finally, let us substitute (\ref{solnTT}) in (\ref{HTT}) and simplify following the procedure of section \ref{s3.1} and steps used to pass from (\ref{H5}) and (\ref{radial}).%
\footnote{We assume that integrals involving $\T^{TT'}_{ab}$ are all well-defined. This is a plausible assumption since $\T_{ab}$ is smooth and of compact support whence its Fourier transform is in Schwartz space.}
We obtain:
\be
E_T \,\,\hat{=}\,\,\f{G}{2\pi}\, \int_{\scrip} {\rm d}T \, {\rm d}^{2}S \,\Big[\partial_{r}\int {\rm d}^{3}x' \T_{ab}^{TT'}(\etar, \vx')\Big]\,
\Big[\partial_{r}\int {\rm d}^{3}x' \T_{cd}^{TT'}(\etar, \vx')\Big]\,
\qo^{ac}\qo^{cd}\, ,\ee
which is manifestly positive.

As we discussed in section \ref{s1}, de Sitter space-time admits gravitational waves whose energy can be arbitrarily negative in the linearized approximation because the time translation Killing field $T^{a}$ is space-like in a neighborhood of $\scrip$.
Indeed, for systems under consideration, gravitational waves satisfy the homogeneous, linearized Einstein's equations in a neighborhood of $\scrip$ and there is an \emph{infinite dimensional subspace of these solutions for which the total energy is negative \cite{abk2}}. What, then, is the physical reason behind the positivity of our $E_{T}$? Consider the shaded triangular region in the left panel of Fig.~\ref{poincare}. It is bounded by $\scrip$,\, upper half of $E^{+}(i^{-})$ and $E^{-}(i^{+})$. The time translation vector field $T^{a}$ is tangential to all these three boundaries, being space-like on $\scrip$, null and \emph{past directed} on the upper half of $E^{+}(i^{-})$, and null and \emph{future directed} on $E^{-}(i^{+})$. As a result, for \emph{any} solution, the energy flux across the upper half of $E^{+}(i^{-})$ is negative, that across $E^{-}(i^{+})$ is positive, and that across $\scrip$ is the sum of the two, which can have either sign and arbitrary magnitude. Thus, the potentially negative energy contribution at $\scrip$ can be traced directly to the incoming gravitational waves across the upper half of $E^{+}(i^{-})$. Now, in the present calculation, physical considerations led us to the \emph{retarded} metric perturbation created by the time varying quadrupoles. Therefore there is no flux of energy across the cosmological horizon $E^{+}(i^{-})$; the potential negative energy flux across $\scrip$ is simply absent. The entire energy flux across $\scrip$ equals the energy flux across $E^{-}(i^{+})$ which is always positive because $T^{a}$ is future directed there. To summarize then, while in general the energy flux across $\scrip$ can have either sign, the metric perturbation $\bar{\chi}_{ab}$ at $\scrip$ created by physically reasonable sources are so constrained that the energy carried by gravitational waves across $\scrip$ is necessarily positive.\\ 
  
(5) The fifth feature concerns time dependence of the source. Equations satisfied by the full (trace-reversed, rescaled) metric perturbation $\bar\chi_{ab}$ refer only to the background metric $\bar{g}_{ab}$ and $T^{a}$ is a Killing field of $\bar{g}_{ab}$ which is time-like in the region in which the source $T_{ab}$ resides. Therefore, it follows that if the source is static, i.e., if $\mathcal{L}_{T} T_{ab} =0$, then the retarded solution $\bar{\chi}_{ab}$ must satisfy $\mathcal{L}_{T}\, \bar\chi_{ab} =0$. Physically, one would expect there to be no flux of energy across $\scrip$. But this is not manifest in Eq.~(\ref{energy}) since it contains a term $\Qp_{ab}$ that does \emph{not} involve a time derivative. Let us therefore examine the fields that enter the definitions (\ref{eq:quadr}) and (\ref{eq:quadp}) of quadrupole moments. A simple calculation shows that, \emph{if $\mathcal{L}_{T}\, T_{ab} =0$}, the fields that enter the definitions of quadrupole moments satisfy
\be \mathcal{L}_{T}\,{\rho}=0;\,\,\,\,\mathcal{L}_{T}\,p=0; \,\,\,\,\mathcal{L}_{T}\, a(\eta) x_{\b{b}}= 0; \,\, \,\,\mathcal{L}_{T}\, \eo^{\b{a}}_{a} = -H\, \eo^{\b{a}}_{a}; \ee
and the 3-dimensional volume element ${\rm d}V$ is preserved under the isometry generated by $T^{a}$. Therefore we have:
\be \label{quadtimeder} \mathcal{L}_{T}\, \Qab \,=\, -2H\, \Qab \quad {\rm and}\quad \mathcal{L}_{T}\,\QPab \,=\, -2H\, \QPab. \ee
Thus, in contrast to what happens in the Minkowski space-time calculation, because of the expansion of the de Sitter scale factor, now $\mathcal{L}_{T}\, T_{ab} =0$ does not imply that quadrupoles are left invariant by the flow generated by $T^{a}$. However, using  (\ref{approxsolnquad}), (\ref{energy}), (\ref{power}) and (\ref{quadtimeder}), it immediately follows that 
\be {\hbox{\rm{if\quad $\mathcal{L}_{T}\, T_{ab} =0$ everywhere,\,\,\, then}}}\quad\quad
E_{T}\, \hat{=}\, 0, \,\, \quad {\rm and}\quad P_{T}(T_{0})\, \hat{=}\, 0 \ee
for all $T_{0}$. (In fact, it follows from Eq. (\ref{power}) that an `instantaneous' result also holds: if\quad $\mathcal{L}_{T}\, T_{ab}\mid_{\eta=\etar} =0$, then $P_{T}(T_{0}) \hat{=} 0$ where $\etar = \eta-r_{0} \equiv -r_{0}$ and $T_{0} = \ln(r_{0}H)$.) Thus, the presence of the term without a time derivative of the pressure quadrupole $\QPab$ is in fact essential to ensure that if $\mathcal{L}_{T}\, T_{ab} =0$ then $E_{T}$ and $P_{T}(T_{0})$ vanish on $\scrip$.%
\footnote{This consistency would have been obscured if we had ignored the pressure terms relative to the density terms in (\ref{conservation}), and used the resulting approximation (\ref{ignorep}) to arrive at the expression of $\chi_{ab}$. That is why we kept all the pressure quadrupole terms even though they can be ignored relative to the analogous density quadrupole terms for Newtonian fluids.\label{note}} \\ 
    
(6) Next, let us consider the limit $\Lambda \to 0$. As discussed in \cite{abk2}, the limit is subtle and has to be taken in the $(t,\vx)$ (rather than the $(\eta, \vx)$) chart. Since the $(t,\vx)$ chart breaks down at $\scrip$ (where $\eta=0$ but $t=\infty$), we cannot directly take the limit of our final expression of the energy flux at $\scrip$ of de Sitter space-time. Rather, we have to `pass through' the physical space-time as in \cite{abk2} and use results from the covariant phase space framework relating expressions involving the $TT$ and $tt$ decompositions in Minkowski space-time. As a result, the procedure is rather long and we will only summarize the main steps here. 

Consider the 1-parameter family of de Sitter backgrounds $\bar{g}_{ab}^{(\Lambda)}$, parametrized by $\Lambda$, with a 1-parameter family $T_{ab}^{(\Lambda)}$ of stress-energy tensors, each satisfying the conservation law with respect to the respective $\bar{g}_{ab}^{(\Lambda)}$ and the condition $\mathcal{L}_{T} T_{ab}^{(\Lambda)} =0$ outside a compact time interval. Let $\chi_{ab}^{(\Lambda)}(t,\vx)$ denote the retarded solutions (\ref{approxsoln}) to the field equations and gauge conditions. For each $\Lambda$, one can express this solution in terms of the source quadrupoles as in (\ref{approxsolnquad}). The question is whether as $\Lambda\to 0$ this 1-parameter family of solutions has a well-defined limit $\mathring\chi_{ab}(t,\vx)$. If so, the analysis in section IV.B.2 of \cite{abk2} shows that: i)  $\mathring\chi_{ab}(t,\vx)$ satisfies the dynamical equation and gauge conditions w.r.t. the Minkowski metric $\etao_{ab}$; and, ii) the expression (\ref{energy}) of energy in the gravitational waves has a well-defined limit, which is furthermore precisely the energy in the solution  $\mathring\chi_{ab}(t,\vx)$, calculated in Minkowski space-time. 

We have already shown in section \ref{s2} that the exact retarded solutions do tend to the exact retarded solution in Minkowski space-time. We will now show that this is also the case for the approximate solutions (\ref{approxsolnquad}). In the $(t, \vx)$ chart, one can perform the integral in the tail term $\flat_{ab} (t,\vx)$ in the solutions (\ref{approxsolnquad})\,  to find that $\flat_{ab} (t,\vx)$ has an explicit overall factor of $H$ whence, as one would expect, the limit $\Lambda \to 0$ of this term vanishes (see Appendix A). Next consider the sharp term $\sharp_{ab}(t,\vx)$ in (\ref{approxsolnquad}). In the $\Lambda\to 0$ limit, we have $T^{a} \to t^{a}$, a time translation in Minkowski metric $\etao_{ab}$;\,\, $\mathcal{L}_{T} \to \mathcal{L}_{t}$;\,\, $a(t) \to 1$ and $Q_{ab}^{(\rho)} \to  \mathring{Q}_{ab}^{(\rho)}$, the mass quadrupole moment constructed from the limiting stress-energy tensor $\mathring{T}_{ab}$ using the Minkowski metric $\etao_{ab}$. Therefore, the limiting solution is given by
\be\label{chio} \lim_{\Lambda\to 0}\, \chi_{ab}^{(\Lambda)} (t,\vx) = \frac{2G}{r}\, \mathcal{L}_{t}\, \mathcal{L}_{t}\, \mathring{Q}_{ab}^{(\rho)} (t_{\rm ret}) =: \mathring{\chi}_{ab}(t,\vx) \ee  
for all $r \gg d(t)$, where $d$ is the physical size of the source with respect to the Minkowski metric $\etao_{ab}$. Now, since by assumption the source is active for a finite time interval, on a $t= {\rm const}$ surface sufficiently in the future, the support of the initial data of $\mathring{\chi}_{ab}(t,\vx)$ is entirely in a region where the approximation holds. Let us consider only the future of this slice. In that space-time region we have a 1-parameter family of solutions $\chi_{ab}^{(\Lambda)}(t,\vx)$ to the source-free equations whose total energy is given by (\ref{energy}) for each $\Lambda>0$. The limit $\mathring{\chi}_{ab}(t,\vx)$ is well-defined, as required. Therefore,  in the $\Lambda \to 0$ limit the energy expression (\ref{energy}) goes over to the energy in $\mathring{\chi}_{ab}(t,\vx)$ with respect to $t^{a}$ in Minkowski space (see Eq (4.24) of \cite{abk2}). And we know that this energy is given by the Einstein formula. Thus, in the limit $\Lambda \to 0$ one recovers the standard quadrupole formula in Minkowski space-time.

To summarize, our energy expression (\ref{energy}) arises as the Hamiltonian on the covariant phase space of linearized solutions on de Sitter space-time, and using results from \cite{abk2} we can conclude that it tends to the expression of the Hamiltonian in Minkowski space in the $\Lambda\to 0$ limit, which in turn reduces to the Einstein flux formula at $\mathring\scri^{+}$. The argument is indirect mainly because in linearized gravity off \emph{Minkowski space-time} we do not know the relation between the $TT$ and $tt$ decompositions. What we know is only the equality between the two expressions of energy, the first evaluated on space-like planes in terms of the $TT$ decomposition and the second, evaluated at $\scrip$ in terms of $tt$. (For definitions of $TT$ and $tt$ fields see the end of section \ref{s4.2}).\\

(7) So far we have focussed on the energy carried by gravitational waves. Let us now discuss the flux of 3-momentum across $\scrip$. The component of the 3-momentum along a space translation $S^{a}{}_{\b{i}}$ is given by \cite{abk2}
\be \label{momentum} 
P_{\,\b{i}}\, \,\hat{=}\,\, \f{1}{16\pi G H}\,\int_{\scrip}\! \rmd^{3}x\,\, \Ecal_{cd}\,\big(\mathcal{L}_{S_{\b{i}}}\,\chi_{ab}\big)\,\qo^{ac}\qo^{bd}\, \ee
We can again use (\ref{Ecal2}) to express $\Ecal_{cd}$ in terms of $\chi_{cd}$: $\Ecal_{ab} = [\f{1}{\eta}\,(\del_\eta^2 - \frac{1}{\eta} \del_\eta)\chi_{ab}]^{TT}$.  
%
%
Now, it is clear from the expression (\ref{approxsoln}) of $\chi_{ab}$ that its dependence on $\vx$ comes entirely from $\etar$. Therefore, $\chi_{ab}$ in invariant under the parity operation $\Pi :\vx \to -\vx $, whence $\f{1}{\eta}\,\big(\del_\eta^2 - \frac{1}{\eta} \del_\eta \big)\chi_{ab}$ is also invariant. Since the operation of taking the $TT$-part refers only to the 3-metric $\qo_{ab}$, it also commutes with $\Pi$. Hence $\Ecal_{ab}$ is even under $\Pi$. The second term, $S^{m}{}_{\b{i}}\,\Do_{m} \chi_{ab}$ is manifestly odd under $\Pi$ since $S^{a}$ is odd but $\chi_{ab}$ is even. Therefore the integral on the right side of (\ref{momentum}) vanishes. Thus, as in the $\Lambda=0$ case, the gravitational waves sourced by a time changing quadrupole do not carry 3-momentum in the post-de Sitter, first post-Newtonian approximation so long as $D_{o} \ll \ell_{\Lambda}$.\\

(8) Finally, let us consider angular momentum. The flux of angular-momentum in the $\b{i}$-direction is given by \cite{abk2}:
\be \label{J1} J_{\b{i}} \,\hat{=}\,\, \f{1}{16\pi G H}\,\int_{\scrip}\! \rmd^{3}x\,\, \Ecal_{cd}\,\big(\mathcal{L}_{R_{\b{i}}}\,\chi_{ab}\big)\,\qo^{ac}\qo^{bd}\, \ee
where $R^{m}{}_{\b{i}}$ is the rotational Killing field in the $\b{i}$-th spatial direction. Now, since the $\vx$-dependence in $\chi_{ab}$ is derived entirely through $\etar$, we have 
\be \mathcal{L}_{R_{\b{i}}}\chi_{ab} = \chi_{mb}\,\Do_{a}R^{m}{}_{\b{i}} + \chi_{am}\,\Do_{b}R^{m}{}_{\b{i}} =  -2\chi_{m(b}\, \epsilono_{a)n}{}^{m}\, \eo^{n}{}_{\b{i}}\, . \ee
Hence,
\be\label{J2} J_{\b{i}} \,\hat{=}\,\, -\f{1}{8\pi G H}\,\int_{\scrip}\! \rmd^{3}x\,\, \Ecal_{cd}\,\big(\epsilono_{am}{}^{n}\,\, \eo^{m}{}_{\b{i}}\,\, \chi_{nb}\big)\, \qo^{ac}\qo^{bd} \ee
Since $\chi_{cd}$ now appears without a derivative in (\ref{J2}), there is a major difference between the calculations of energy and 3-momentum fluxes across $\scrip$: Now the integral over $\eta'$ in the tail term $\flat_{ab}$ in the expression (\ref{approxsolnquad}) of $\chi_{ab}$ persists. To evaluate the right side of (\ref{J2}), for $\chi_{ab}$ we simplify the tail term $\flat_{ab}$ in \eqref{approxsoln} by carrying out the integral over $\eta'$ (see Appendix), and for $\Ecal_{ab}$ we use Eqs (\ref{Ecal2}) and (\ref{source2}) as in the calculation of the energy flux. These simplifications lead to:
\be \label{J3} J_{\b{i}} \,\hat{=}\, \f{G}{4\pi}\, \int_{\scrip}\!\! {\rm d}T\, {\rm d}^{2}S\,\, \big[\mathcal{R}^{ab}\big]\,\,\, \big[\epsilono_{am}{}^{n}\,\, \eo^{m}{}_{\b{i}}\,\, \left( \ddot{Q}_{nb}^{(\rho)} + H \dot{Q}_{nb}^{(\rho)} +H \dot{Q}^{(p)}_{nb} + H^{2} Q_{nb}^{(p)} \right)\big]^{TT}\, ,  \ee
where, as before $T$ is the affine parameter along the integral curves of the `time translation' Killing field $T^{a}$ and $\mathcal{R}^{ab}$ is defined in (\ref{R}). Note that if the stress-energy satisfies $\mathcal{L}_{T} T_{ab} =0$ at some time $\eta=\eta_{o}$ then the `radiation field' $\mathcal{R}_{ab}$ vanishes on the cross-section $r= \eta_{o}$ on $\scrip$, whence the flux of (energy and) angular momentum vanish on that cross-section. Similarly if $\mathcal{L}_{R_{\b{i}}}\, T_{ab}$ vanishes at $\eta= \eta_{o}$, then the flux of angular momentum vanishes on the cross-section $r=\eta_{o}$. Finally, in the limit $\Lambda \to 0$, using the same argument as that used for energy, one can show that (\ref{J2}) reduces to the standard formula in Minkowski space-time. Again the argument is indirect because the expression of the Hamiltonian generating rotations on the covariant phase space in Minkowski space-time involves the $TT$ part of the solution while the standard expression of angular momentum at null infinity involves the $tt$-part and the explicit relation between the two is not yet known.

%



\section{Discussion}
\label{s5}

Einstein's quadrupole formula has played a seminal role in the study of gravitational waves emitted by astrophysical sources. His analysis was carried out only to the leading post-Newtonian order, assuming that the time-changing  quadrupole is a first order, external source in Minkowski space-time. In spite of these restrictions, his quadrupole formula sufficed to bring to forefront the extreme difficulty of detecting these waves. However, thanks to the richness of our physical universe and ingenuity of observers, impressive advances have occurred over the last four decades.  First, the careful monitoring of the Hulse-Taylor pulsar has provided clear evidence for the validity of the quadrupole formula to a $10^{-3}$ level accuracy. Furthermore gravitational wave observatories, equipped with detectors with unprecedented sensitivity, have led us to the threshold of the era of gravitational wave astronomy. Therefore it is now all the more important that our theoretical understanding of gravitational waves be sufficiently deep to do full justice to the impressive status of the field on the observational front. The goal of this series of papers is to fill a key conceptual gap that still remains: incorporation of the positive cosmological constant in our understanding of the properties of gravitational waves and dynamics of their sources.

Since the observed value of the cosmological constant is so small, one's first reaction is just to ignore its presence. However, as we discussed in section \ref{s1}, even a tiny cosmological constant can cast a long shadow because it abruptly changes the conceptual setup that is used to analyze gravitational waves. As a result, the limit $\Lambda \to 0$ is not necessarily continuous; indeed, some physical quantities --such as the lower bound of the energy carried by gravitational waves-- can be \emph{infinitely} discontinuous. Therefore, without a systematic analysis, one can not be confident that the quadrupole formula would continue to be valid in presence of a positive cosmological constant.\vskip0.15cm

Indeed, our analysis revealed that the presence of a cosmological constant does modify Einstein's analysis in unforeseen ways. In particular:\\
%
%
\emph{(i)} The propagation equation for metric perturbations
in the transverse-traceless gauge is not the wave equation as in Minkowski space-time, but has an effective  mass term (see (\ref{eom})). Although this mass is tiny, there is potential for the differences from Minkowskian propagation to accumulate over cosmological distances to produce $\ord(1)$ departures in the value of the metric perturbation in the asymptotic region;\\ 
\emph{(ii)} The retarded field does not propagate sharply along the null cone of the de Sitter metric. Although the de Sitter metric is conformally flat, since the equation satisfied by the metric perturbation is not conformally invariant, its expression acquires a tail term due to the back-scattering by de Sitter curvature.
%
%
As shown in the Appendix, even in the asymptotic region, the cumulative effects make the tail term comparable to the sharp term (which has the same form as in Minkowski space-time); \\
\emph{(iii)} Since the radial $r$ coordinate goes to infinity $\mathring\scri^{+}$ of Minkowski space-time, the analysis of waves makes heavy use of $1/r$ expansions. These can no longer be used  in de Sitter space-time because $r$ ranges over the entire positive real axis on de Sitter $\scrip$. In  particular, the $tt$-decomposition, that is local in space being tailored to the $1/r$ expansions in Minkowski space-time, is no longer meaningful near de Sitter $\scrip$.\\
\emph{(iv)} The retarded, first order metric perturbation depends not only on the mass quadrupole as in Einstein's calculation but also on the pressure quadrupole. Also, while only the third time derivative of the mass quadrupole features in Einstein's calculation, now we also have a contribution from lower time derivatives of the two quadrupoles, as well as the pressure quadrupole itself; \\ 
\emph{(v)}  The \emph{physical} wavelengths $\lambda_{\rm phys}$ of perturbations grow exponentially as the wave propagates and vastly exceed the curvature radius $\ell_{\Lambda} = H^{-1} \equiv \sqrt{3/\Lambda}$ in the asymptotic region near $\scrip$. Therefore, the geometric optics approximation often used to study the effect of background curvature on propagation of gravitational waves \cite{thorne} fails even for waves produced by `tame sources' such as a compact binary. Since waves `experience' the curvature, their propagation is quite different from that in flat space. Also, since the expression (\ref{H5}) involves the metric perturbation evaluated in the zone where $\lambda_{\rm phys} > \ell_{\Lambda}$, a priori the effect of $\Lambda$ on radiated energy could be non-negligible; \\ 
\emph{(vi)} $\scrip$, the arena used to analyze properties of gravitational waves unambiguously changes its character from being a null future boundary of space-time to a space-like one. As a result, all Killing fields of the background de Sitter space-time --including the `time translation' used to define energy-- are space-like in a neighborhood of $\scrip$. Consequently, while linearized gravitational waves carry positive energy in Minkowski space-time, de Sitter space-time admits gravitational waves carrying arbitrarily large negative energy.\vskip0.15cm

These differences are sufficiently striking to cast a doubt on one's initial intuition that the cosmological constant will have no role in the study of compact binaries. For example, they open up the possibility that Einstein's quadrupole formula could receive significant corrections --e.g., of the order $\ord(H\lambda_{\rm phy})$-- even though the observed value of $H$ is so small. Interestingly, the final expression (\ref{power}) of radiated power shows that this does not happen for astrophysical processes such as the Hulse-Taylor binary pulsar, or the compact binary mergers that are of greatest interest to the current ground based gravitational wave observatories. How does this come about? Why do the qualitative differences noted in the last paragraph not matter in the final result for these systems? The physical reasons can be summarized as follows: \\ 
\emph{(a)} First, while the propagation of $\chi_{ab}$ is indeed not sharp, what matters for radiated energy are certain \emph{derivatives} of $\chi_{ab}$ and these do have sharp propagation.\\ 
\emph{(b)} Second, while the final expressions (\ref{energy}) and (\ref{J3}) of radiated energy and angular momentum are evaluated at $\scrip$, the integrand refers to the time derivatives of quadrupole moments evaluated at retarded instants of time. In our $c=1$ units, even though (\ref{energy}) and (\ref{J3}) involve fields at late times, the time scales in the `dots' in these expressions are determined by $\lambda_{\rm phy}^{\rm source}$, the wave length evaluated at the source, and \emph{not} by the exponentially larger physical wavelengths $\lambda_{\rm phy}^{\rm asym}$ \,in the asymptotic region. Therefore for the sources on which gravitational wave observatories will focus in the foreseeable future, $H\ddot{Q}_{ab}^{(\rho)}$, for example, is suppressed  relative to $\dddot{Q}_{ab}^{(\rho)}$ by the factor $H\,\lambda_{\rm phy}^{\rm source}$ (rather than enhanced by the factor  $H\,\lambda_{\rm phy}^{\rm asym}$) and  $\dddot{Q}_{ab}^{(\rho)}$ completely dominates over the remaining 5 terms (which have $H,H^{2}$ or $H^{3}$ as coefficient). In particular, the pressure quadrupole can be neglected for these sources. Had our expression of power referred to time scales associated with the asymptotic values of $\lambda_{\rm phy}$, effects discussed in the previous paragraph would have completely altered the picture. Then, the terms with the highest powers of $H$ --in particular the pressure quadrupole $\Qp_{ab}$ term--  would have dominated and the contribution due to $\dddot{Q}_{ab}^{(\rho)}$ would have been completely negligible! \\
\emph{(c)} Third, while a neighborhood of $\scrip$ in the Poincar\'e patch $(\man, \bar{g}_{ab})$ does admit gravitational waves carrying arbitrarily large negative energies, our calculation showed that such waves can not result from time-changing quadrupoles. The reason is simplest to explain using the shaded region in the left panel of Fig.~\ref{poincare}. Negative contribution to the energy at $\scrip$ can come only from the waves that arrive from the upper half of $E^{+}(i^{-})$. But the physics of the problem led us to consider \emph{retarded} solutions with the given $T_{ab}$ as source and for these solutions there is no energy flux at all across  $E^{+}(i^{-})$. This is why our energy flux (\ref{energy}) across $\scrip$ is necessarily positive.\vskip0.15cm

Because of these reasons, for binary coalescences that are of greatest interest to the current gravitational wave observatories, energy and power are determined essentially by the third time derivative of the mass quadrupole, as in Einstein's formula. This quadrupole moment (\ref{eq:quadr}) is calculated using the physical de Sitter geometry and the time derivative `overdot' refers to the Lie derivative with respect to the de Sitter time translation $T^{a}$ specified in (\ref{T}). However, in the limit $\Lambda \to 0$, it goes over the mass-quadrupole used in Einstein's formula. Therefore, for compact binaries of interest to the current gravitational wave observatories, the difference is again negligible.

However, there are some circumstances in  which the differences between the $\Lambda=0$ and $\Lambda >0$ could be significant. First, consider the tail term in the expression (\ref{approxsolnquad}) of $\chi_{ab}$.  Since it arises because of back-scattering due to de Sitter curvature, it is proportional to $H$. However, it involves an integral over a cosmologically large time interval which could compensate the smallness of $H$ and make the tail term comparable to the one that arises from sharp propagation. The tail term could then yield a significant new contribution to the memory effect \cite{memory1,memory2,memory3} for detectors placed near $\scrip$. A second example is provided by mergers of supermassive black holes at the centers of two different galaxies, such as Milky way and Andromeda. Since the time scales associated with such galactic coalescences are cosmological, the various effects discussed above will come into play. Gravitational waves created in this process will have \emph{extremely} long wavelength already at inception, making the departures from Einstein's quadrupole formula significant. While these waves will not be detected directly in any foreseeable future, they provide a background which could have indirect influences. An illustration of this general mechanism is provided by inflationary cosmology, where super-horizon modes can induce non-Gaussiantities in observable modes due to mode-mode coupling resulting from non-linearities of general relativity (see, e.g., \cite{sh,ia}). 

To conclude, we note that this analysis also provides some hints for the gravitational radiation theory in full, non-linear general relativity with a positive $\Lambda$ which would be of interest to geometric analysis, because of issues such as the positivity of total energy. First, to describe an isolated gravitating systems such as an oscillating star, or one collapsing to  form a black hole, or a compact binary, it would be appropriate to consider only the portion of full space-time that is bounded in the future by $\scrip$ and in the past by the future cosmological event horizon $E^{+}(i^{-})$, where the point $i^{-}$ represents the past time-like infinity defined by the source. This is because the isolated system and the radiation it emits would be invisible to the rest of the space-time. Second, the `no-incoming radiation' boundary condition will have to be imposed on the past boundary, $E^{+}(i^{-})$. Since this is an event horizon, a natural strategy would be to demand that it be a weakly isolated horizon \cite{wih1,wih2,akrev}. 
It would be interesting to analyze if this condition would suffice to ensure that the flux of energy across $\scrip$ is positive, as in the weak field limit discussed here. If so, one would have the desired generalization of the celebrated result due to Bondi and Sachs that gravitational waves carry away \emph{positive} energy, in spite of the fact that the corresponding  asymptotic `time translation' on $\scrip$ would now be space-like for $\Lambda >0$. Third, results of \cite{abk1} and \cite{abk2} suggest that there will be a 2-sphere `charge integral' --representing the generalization of the notion of Bondi-Sachs energy to the $\Lambda>0$ case-- and the difference between charges  associated with two different 2-spheres will equal the energy flux across the region bounded by the two 2-spheres. A natural question is whether this charge is also positive.%
\footnote{These Bondi-type charge integrals will also refer to an asymptotic `time-translation'. They will be distinct from the ADM-type charge-integral associated with a \emph{conformal} --rather than time-translation-- symmetry discussed in \cite{conformal-energy}, and the intriguing 2-sphere integral recently discovered \cite{st}, both of which are known to be positive.} 
Fourth, the form (\ref{approxsolnquad}) of the solution $\chi_{ab}$ at $\scrip$ implies that the recently proposed \cite{hc} generalization of Bondi-type expansions for full general relativity can describe at most half the desired set of asymptotically de Sitter space-times. A further generalization is necessary to capture both polarizations at $\scrip$. Finally, in the linear approximation considered in this paper, the past cosmological event horizon $E^{-}(i^{+})$ of the point at future time-like infinity could be taken to lie in the `far zone'. Furthermore, since there is no incoming radiation across $E^{+}(i^{-})$ from (the shaded portion of the left panel of) Fig.~\ref{poincare} it follows that the flux of energy across $E^{-}(i^{+})$ equals that across $\scrip$ and is, in particular, positive. In  full, non-linear general relativity, then, $E^{-}(i^{+})$ may well serve as an `approximate' $\scrip$ to analyze gravitational waves. Because this surface is null, it may be easier to compare results in the $\Lambda >0$ case with those in the $\Lambda=0$ case in full general relativity.

\section*{Acknowledgment} AA thanks Thibault Damour, Ghanshyam Date, Xiao Zhang and especially Eric Poisson for discussions. AK and BB thank Leo Stein for discussions and drawing attention to some references. Thanks are also due to an anonymous referee for bringing to our notice Ref 21. This work was supported in part by the NSF grants PHY-1205388 and PHY-1505411, the Eberly research funds of Penn State and Frymoyer Fellowships to AK and BB.\goodbreak

\begin{appendix}
\section{The tail term}
\label{a1}

A qualitative difference between the $\Lambda>0$ and $\Lambda=0$ cases is the presence of the tail term in the retarded solution. 
In this appendix we will discuss some properties of this term.
The first natural question is whether it disappears in the $\Lambda \to 0$ limit, i.e., whether the limit is continuous. The second conceptually important question is whether $\flat_{ab}$ is negligible compared to the sharp term $\sharp_{ab}$ if $\Lambda \not=0$ but tiny. We will now show that the answer to the first question is in the affirmative but that to the second question is in the negative. This is another illustration of the subtlety of the limit $\Lambda \to 0$.

To answer these questions, it is most convenient to work in the $(t,\vx)$ chart. Now the tail term assumes the form
\be \label{tail1} \flat_{ab}(t,\vx)\, = \,-2GH \int_{-\infty}^{t_{\rm ret}}\!\!\! {\rm d}t' \Big[\dddot{Q}^{(\rho)}_{ab} + 3H\ddot{Q}^{\rho}_{ab} + 2H^{2} \dot{Q}^{(\rho)}_{ab} + H \ddot{Q}^{(p)}_{ab} + 3H^{2}\dot{Q}^{(p)}_{ab} + 2H^{3} Q^{(p)}_{ab} \Big](t')\, . \ee
In the $\Lambda\to 0$ limit, the `overdot' tends to the well-defined Lie derivative with respect to a time translation Killing vector field in Minkowski space-time. Therefore the overall multiplicative factor $H$ in (\ref{tail1}) makes it transparent that  $\flat_{ab}$ does vanish in the $\Lambda\to 0$ limit. 

To answer the second question, let us use the fact that $\dot{Q}_{ab} = \partial_{t}Q_{ab} - 2H Q_{ab}$ to carry out the integral over $t$ in (\ref{tail1}). Then, we have:
\be \flat_{ab}(t, \vx)\, = \,-2GH\,\,  \Big[\,\ddot{Q}_{ab}^{(\rho)} + H \dot{Q}_{ab}^{(\rho)} + H \dot{Q}_{ab}^{(p)} + H^{2}Q_{ab}^{(p)}\,\Big]_{-\infty}^{t_{\rm ret}}\, . \ee
As shown in section \ref{s4.3}, the assumption $\mathcal{L}_{T}\, T_{ab} =0$ in the distant past implies $\dot{Q}_{ab}^{(\rho)} = -2 H Q_{ab}^{(\rho)}$ there (and similarly for the pressure quadrupole). Therefore, we have
\be \flat_{ab}(t,\vx)\, = -2GH\,\big[\,\ddot{Q}_{ab}^{(\rho)}+ H \dot{Q}_{ab}^{(\rho)}+ H \dot{Q}_{ab}^{(p)} + H^{2}Q_{ab}^{(p)}\,\Big](t_{\rm ret}) \,+\, 2GH^{3}\,C_{ab}\, , \ee
%
%
where $C_{ab}$ is just a constant term. It does not play any role in the calculation of energy flux because in the expression (\ref{H5}) only derivatives of $\chi_{ab}$ appear. In the expression (\ref{J1}) of the flux of angular momentum, $\chi_{ab}$ does appear without 
a derivative but the constant term does not contribute  because it is integrated against $\Ecal^{ab}$ which is of compact support and divergence-free on $\scrip$. Finally, since it is constant, it will not feature in the analysis of the memory effect as well.

With this simplification of the tail term, we can return to (\ref{approxsolnquad}) and, for $r>-\eta$, write $\chi_{ab}$ as
%
%
\be \chi_{ab}(\eta,\vx) = \f{2G}{R(\etar)}\,\,\big[(1- \f{r}{r-\eta})
\,\ddot{Q}_{ab}^{(\rho)}\big]\, + \ord(H) \ee
where $R(\etar) = r a(\etar)$ is the physical distance between the source and the point $\vx$ at time $\eta=\etar$ (and terms $\ord(H)$ vanish in the limit $\Lambda \to 0$). The factor $1$ in the square bracket  comes from the sharp term while the factor $r/(r-\eta)$ comes from the tail term. At late times the two contributions are comparable and at $\scrip$ they are in fact equal in magnitude but opposite in sign. This occurs no matter how small $\Lambda$ is! The remainder --i.e., the $\ord{H}$ term-- at $\scrip$ has contributions from both the sharp and the tail terms:
\be \chi_{ab}(\vx) \hat{=}\, 2GH^{2}\,[\dot{Q}^{(\rho)}_{ab} + H Q_{ab}^{(p)}]\, + 2H^{3} C_{ab}\, .\ee

This analysis provides the precise sense in which the back-scattering effects encoded in the tail term --which can also be thought of as arising from the addition of a mass term to the propagation equation of $\bar\gamma_{ab}$-- provide an $\ord(1)$ contribution to the metric perturbation $\chi_{ab}$ near $\scrip$. This is a concrete realization of the non-trivial outcome of the secular accumulation of small effects we referred to in section \ref{s1}. Finally, as mentioned after Eq. (\ref{source}), the tail term is essential to make the field $\Ecal_{ab}$ finite at $\scrip$. As a result, it contributes on an equal footing as the sharp term to the expression of energy and angular momentum radiated across $\scrip$.

\end{appendix}


\end{document}